\documentclass[12pt]{iopart}
\usepackage{iopams}
\usepackage{graphicx}
\expandafter\let\csname equation*\endcsname\relax
\expandafter\let\csname endequation*\endcsname\relax
\usepackage{amsmath}
\usepackage{natbib}

\begin{document}

\title[The three-body problem]
{The three-body problem}

\author{Z.E. Musielak$^1$ and B. Quarles$^2$} 

\address{$^1$Department of Physics, The University of Texas 
at Arlington, Arlington, TX 76019, USA\\
$^2$Space Science and Astrobiology Division 245-3, NASA Ames 
Research Center, Moffett  Field, CA 94035, USA}

\ead{zmusielak@uta.edu; billy.l.quarles@nasa.gov}

\begin{abstract}
The three-body problem, which describes three masses interacting 
through Newtonian gravity without any restrictions imposed on the 
initial positions and velocities of these masses, has attracted 
the attention of many scientists for more than 300 years.  In this 
paper, we present a review of the three-body problem in the context 
of both historical and modern developments.  We describe the general 
and restricted (circular and elliptic) three-body problems, different 
analytical and numerical methods of finding solutions, methods for performing 
stability analysis, search for periodic orbits and resonances, 
and application of the results to some interesting astronomical and space 
dynamical settings.  We also provide a brief presentation of the 
general and restricted relativistic three-body problems, and discuss 
their astronomical applications. 
\end{abstract}

\maketitle

\section{Introduction} 

In the three-body problem, three bodies move in space under their 
mutual gravitational interactions as described by Newton's theory 
of gravity.  Solutions of this problem require that future and past 
motions of the bodies be uniquely determined based solely on their 
present positions and velocities.  In general, the motions of the 
bodies take place in three dimensions (3D), and there are no 
restrictions on their masses nor on the initial conditions.  Thus, 
we refer to this as the {\it general three-body problem}.  At first 
glance, the difficulty of the problem is not obvious, especially when 
considering that the two-body problem has well-known closed form 
solutions given in terms of elementary functions.  Adding one extra 
body makes the problem too complicated to obtain similar types of 
solutions.  In the past, many physicists, astronomers and mathematicians 
attempted unsuccessfully to find closed form solutions to the three-body 
problem.  Such solutions do not exist because motions of the three bodies 
are in general unpredictable, which makes the three-body problem one of 
the most challenging problems in the history of science.

In celestial mechanics, the general three-body problem deals with 
gravitationally interacting astronomical bodies and intends to predict 
their motions.  In our Solar System, the planets and asteroids move 
around the Sun, while the moons orbit their host planets, which in 
turn also move around the Sun.  As typical examples of the three-body 
problem, we may consider the Sun-planet-planet, Sun-planet-moon, 
or Sun-planet-asteroid systems.  The three-body problem 
representing the latter system can be significantly simplified because 
the mass of the asteroid is always negligible when compared to the mass 
of either the Sun or the planet, which means that the gravitational 
influence of the asteroid on the planet and the Sun can be omitted 
from the theory.  If this condition is satisfied, then the general 
three-body problem becomes the {\it restricted three-body problem}, 
and there are two possibilities, namely, the two bodies with dominant 
masses move around their center of mass either along  circular or elliptic 
orbits, which leads to the respective {\it circular} or {\it elliptic 
restricted three-body problems}.  It is the circular restricted 
three-body problem (CR3BP) that has been the most extensively studied.

The three-body problem has been studied for over three hundred years.
We wish to provide a brief historical overview of the most significant 
results that have contributed to the current developments in the field.
We divide this overview into two parts: first, we highlight the time 
period from the publication of Newton's {\it Principia} in 1687
to the publication of Poincar\'e's {\it Les M\'ethodes Nouvelles de la 
M\'ecanique C\'eleste} published in 1892 (volume 1 and 2) and in 1899 
(volume 3); in the second part we focus on developments that took place 
from 1900 to the present time.

\subsection{From Newton to Poincar\'e}

The three-body problem was formulated and studied by \cite{Newton1687} 
in his {\it Principia}, where he considered the motion of the Earth and 
the Moon around the Sun.  In this problem there are key elements such as 
the ratio of the Moon's mass, $M_{\rm M}$, to the Earth's mass, $M_{\rm E}$, 
that is $M_{\rm M} / M_{\rm E} = 0.0123$, which is a small but not negligible number.  Another element is the tilt of the Moon's orbit to the Earth's 
orbit, about $5^{\circ}$.  This makes the problem completely predictable 
despite the fact that the Earth's orbit around the Sun and the Moon's 
orbit around the Earth are almost perfect circles.  Newton stated that 
the problem was very difficult to solve; however, he was able to obtain 
an approximate solution which agreed with observations to within $8\%$.  

A special form of the general three-body problem was proposed by 
\cite{Euler1767}.  He considered three bodies of arbitrary (finite) masses 
and placed them along a straight line.  Euler showed that the bodies would 
always stay on this line for suitable initial conditions, and that the line 
would rotate about the center of mass of the bodies, leading to periodic 
motions of all three bodies along ellipses.   Around the same time, \cite
{Lagrange1772} found a second class of periodic orbits in the general 
three-body problem.  He showed that if the bodies were positioned in such a way 
that they form a triangle of equal sides which would move along ellipses for 
certain initial conditions, preserving always their original configuration.  
The Euler and Lagrange solutions are now known as particular solutions 
to the general three-body problem, and they will be further discussed 
in Section 3.

\cite{Euler1767} was the first to formulate the CR3BP in a rotating (or 
synodic) coordinate system.  This was an important development in the 
study of the three-body problem.  Lagrange also studied the CR3BP and 
demonstrated that there were five equilibrium points (now known as the 
Lagrange points) at which the gravitational forces of the bodies balanced 
out; the Trojan asteroids discovered in 1906 along Jupiter's orbit occupy 
space close to two of the Lagrange points.  

Important contributions to the CR3BP were made by \cite{Jacobi1836}, 
who used the synodic (rotating) coordinate system originally introduced 
by \cite{Euler1767} to demonstrate that there was an integral of motion, 
which is now named after him.  The Jacobi integral was used by 
\cite{Hill1877,Hill1878} to determine the motion of an asteroid in the 
three-body problem and to introduce the so-called zero velocity curves 
(ZVC), which establish regions in space where the bodies are allowed to 
move.  Hill considered a special case of the CR3BP in which two masses 
were much smaller than the first one (the problem is now known as the 
Hill problem), and in this way he discovered a new class of periodic 
solutions.  His main contribution was to present a new approach to 
solve the Sun-Earth-Moon three-body problem.  After almost two hundred 
years since the original formulation of the problem by \cite{Newton1687}, 
Hill developed his lunar theory, which with some modifications made by 
\cite{Brown1896}, is still being used today in celestial mechanics 
\citep{Gutzwiller1998}.

In the second half of the nineteenth century, Poincar\'e studied and 
advanced the solution of the three-body problem.  His monumental 
three-volume book {\it Les M\'ethodes Nouvelles de la M\'ecanique C\'eleste}, originally published in 1892-99, contains his most important contribution 
to the study of the CR3BP; the book was translated as {\it New Methods of 
Celestial Mechanics} and edited by L. D. Goroff in 1993.  In this work, 
Poincar\'e developed a number of new qualitative methods to solve differential equations and used them to identify and study possible periodic orbits, while demonstrating non-integrability of the system of equations describing the 
three-body problem.  Poincar\'e's new methods allowed him to identify the 
unpredictability of the problem, and to discover the first manifestation 
of a new phenomenon, which is now commonly known as chaos.  Poincar\'e 
submitted some of those results for the King Oscar II Birthday Competition
and he was awarded the prize.  The competition and Poincar\'e's other 
contributions are described by \cite{BarrowGreen1997}, \cite{Diacu1996} 
and \cite{Peterson1993}, and will not be repeated here.

\subsection{From Poincar\'e to the present time}

\cite{Poincare1892} studied Hill's problem and generalized Hill's definition 
of periodic orbits \citep{Hill1877,Hill1878}.  He was able to choose such 
initial conditions that resulted in periodic orbits in the special {\it 
restricted} three-body problem.  Poincar\'e's work on the existence of 
periodic solutions in dynamical systems with one degree of freedom was 
extended by \cite{Bendixson1901}.  He formulated and proved a theorem that 
is now known as the Poincar\'e-Bendixson theorem, which gives a criterion 
for the existence of periodic solutions in such systems.  The work of Poincar\'e 
led to systematic searches for periodic orbits in the three-body problem and 
their classification by \cite{Darwin1897,Darwin1909}, \cite{Moulton1920} and 
\cite{Stromgren1922}, as well as extensive studies of stability of such orbits, 
which was initiated by \cite{Poincare1892} then continued by \cite{LeviCivita1901} 
and \cite{Lyapunov1907}, who significantly generalized Poincar\'e's approach and 
his results.  

Further generalizations and extensions of Poincar\'e's ideas on the stability 
of periodic orbits were pursued by \cite{Birkhoff1912}.  He introduced the 
concept of recurrent motion, which requires a sufficiently long time so that 
the motion comes arbitrarily close to all its states of motion, and showed its relationship to orbital stability.  In addition, \cite{Birkhoff1913} proved 
the `Last Geometric Theorem' formulated by Poincar\'e.  The Poincar\'e-Birkhoff theorem states that there are infinitely many periodic orbits near any stable periodic orbit; it also implies the existence of quasi-periodic orbits.  
\cite{Birkhoff1915} also developed a topological model for the restricted 
three-body problem in which an asteroid was confined to move inside an oval 
about one of the larger masses. 

By searching for periodic and non-periodic solutions to the three-body problem, 
researchers realized that the differential equations describing the problem 
contain singularities at which the solutions are abruptly terminated.  An 
example of such a termination are collisions between two or even all three 
bodies; thus, we have the so-called collision singularities.  Once their 
presence is established, the next challenging task is to eliminate them by 
a process called regularization.  An important work on singularities in the 
three-body problem was done by \cite{Painleve1896,Painleve1897}.  He 
determined that collisions were the only singularities and that collisions 
can be excluded by setting certain initial conditions for which the equations 
of motion of the three-body problem could be integrable using power series solutions.  Painlev\'e did not find such solutions but his work had a strong 
impact on other researchers in the field.  The problem of the regularization of collisions between two bodies was investigated mainly by \cite{LeviCivita1903}, 
\cite{Bisconcini1906}, and \cite{Sundman1907, Sundman1909,Sundman1912}.  They 
also studied the triple collisions and formulated theorems that allowed 
establishing conditions for such collisions \citep{Siegel1991}.

Realizing that the possibility of finding closed-form solutions to the 
three-body problem was unlikely, researchers considered finding infinite 
series solutions.  \cite{Delaunay1867} used canonical variables in the 
perturbation theory of the three-body problem.  Following Delaunay's work, 
\cite{Lindstedt1884} and \cite{Gylden1893} introduced infinite series and investigated the rates of their convergence, but they were not able to 
present a formal proof of the convergence.  \cite{Poincare1892} clarified 
the concepts of convergence and divergence of the infinite series and exposed certain flaws in the methods developed by these authors.  \cite{Painleve1896,
Painleve1897} also attempted to find solutions to the three-body problem 
given in terms of infinite series but failed as the others did before him.  However, he clearly stated that such solutions were possible, at least in principle.      
  
Indeed, Painlev\'e was right and \cite{Sundman1912} found a complete solution 
to the three-body problem given in terms of a power series expansion.  
Unfortunately, Sundman's solution converges very slowly so that it cannot be 
used for any practical applications.  An important question to ask is: Does 
the existence of Sundman's solution contradict the unpredictability of motions 
in the three-body problem postulated by \cite{Poincare1892}?  Well, the answer 
is that it does not because trajectories of any of the three bodies cannot be determined directly from Sundman's solution, which means that the trajectories 
can still be fully unpredictable, as this is observed in numerical simulations 
(see Section 4, 5 and 8). 

The three-body system considered here is a Hamiltonian system, which means that 
its total energy is conserved.  \cite{Poincare1892} expressed the differential equations describing the three-body problem in the Hamiltonian form and discussed 
the integrals of motion \citep{BarrowGreen1997}.  In general, Hamiltonian systems 
can be divided into integrable and non-integrable.  As shown above, even the 
systems that are in principle non-integrable may still have periodic solutions 
(orbits) depending on a set of initial conditions.  Poincar\'e, Birkhoff and 
others considered quasi-periodic solutions (orbits) in such systems.  The very fundamental question they tried to answer was: What happens to solutions of an integrable Hamiltonian system when the governing equations are slightly perturbed?  The correct answer to this question was first given by \cite{Kolmogorov1954} but 
without a formal proof, which was later supplied independently by \cite{Moser1962}
and \cite{Arnold1963}; the formal theorem is now known as the Kolmogorov-Arnold-Moser (KAM) theorem.  This theorem plays an important role in the three-body problem and in other Hamiltonian dynamical systems \citep{Hilborn1994}.
  
As mentioned above, periodic orbits in the three-body problem were discovered 
in the past by using analytical methods.  By utilizing computer simulations, 
\cite{Henon1965,Henon1974} and \cite{Szebehely1967} found many periodic orbits 
and classified them.  Recent work by \cite{Suvakov2013a} shows that periodic orbits
are still being discovered.  An interesting problem was investigated numerically 
by \cite{Szebehely1967a,Szebehely1967b}, who considered three objects with their 
masses proportional to 3, 4, and 5 located at the vertices of a Pythagorean 
triangle with the sides equal to 3, 4, and 5 length units; they were able to 
show that two of these objects form a binary system but the third one was 
expelled at high speed from the system.  More modern discoveries include 
a new 8-type periodic orbit \citep{Moore1993,Chenciner2000,Suvakov2013a} 
and 13 new periodic solutions for the general three-body problem with 
equal masses \citep{Suvakov2013a}. 

Modern applications of the three-body problem have been greatly extended
to include the Earth, Moon, and artificial satellites, as well as the 
recently discovered distant extrasolar planetary systems (exoplanets).  
With the recent progress in detection techniques and instrumentation, 
there are over a thousand confirmed exoplanets orbiting single 
stars, either both or a single component within binary stars, 
and even triple stellar 
systems{\footnote{http://exoplanet.eu/catalog/}} along with thousands 
of exoplanet candidates identified by the NASA's {\it Kepler} space 
telescope{\footnote{http://kepler.nasa.gov/mission/discoveries/candidates/}}.  
A typical three-body system would be a single star hosting two exoplanets 
or a binary stellar system hosting one exoplanet; we discuss such systems 
in this paper. 
  
The three-body problem is described in many celestial mechanics books
such as \cite{Whittaker1937}, \cite{Wintner1941}, \cite{Pollard1966}, 
\cite{Danby1988}, \cite{Siegel1991}, \cite{Murray1999}, \cite{Roy2005} 
and \cite{Beutler2005}, and in review papers such as \cite{Holmes1990}, 
\cite{Szebehely1997}, \cite{Gutzwiller1998} and \cite{Ito2007}.  The 
books by \cite{Poincare1892}, \cite{Szebehely1967}, \cite{Marchal1990}, 
\cite{BarrowGreen1997} and \cite{Valtonen2006} are devoted exclusively 
to the three-body problem and also discuss its many applications.  

This paper is organized as follows: in Section 2 we formulate and discuss 
the general three-body problem; then we describe analytical and numerical 
studies of this problem in Sections 3 and 4, respectively; applications of 
the general three-body problem are presented and discussed in Section 5; 
the circular and elliptic restricted three-body are formulated and discussed 
in Sections 6 and 7, respectively; applications of the restricted three-body problem are given in Section 8; the relativistic three-body problem and its astrophysical applications are described in Section 9; and we end with a 
summary and concluding remarks in Section 10.

\section{The general three-body problem}

\subsection{Basic formulation}

In the general three-body problem, three bodies of arbitrary masses 
move in a three dimensional (3D) space under their mutual gravitational 
interactions; however, if the motions of the bodies are confined to one 
plane the problem is called the {\it planar general three-body problem}.  
Throughout this paper, we use Newton's theory of gravity to describe the gravitational interactions between the three bodies; the only exception 
is Section 9 in which we discuss the relativistic three-body problem.   

To fully solve the general three-body problem, it is required for the 
past and future motions of the bodies to be uniquely determined by their 
present positions and velocities.  Let us denote the three masses by $M_i$, 
where $i$ = 1, 2 and 3, and their positions with respect to the origin of 
an inertial Cartesian coordinate system by the vectors $\vec R_i$, and 
define the position of one body with respect to another by $\vec r_{ij} = 
\vec R_j - \vec R_i$, where  $\vec r_{ij} = - \vec r_{ji}$, $j$ = 1, 2, 3 
and $i \neq j$.  With Newton's gravitational force being the only force 
acting upon the bodies, the resulting equations of motion are    
\begin{equation}
M_i {{d^2 \vec R_i} \over {dt^2}} = G \sum_{j=1}^{3} {{M_i M_j} 
\over {r_{ij}^3}} \vec r_{ij}\ ,
\label{S2eq1}
\end{equation}

\noindent
where $G$ is the universal gravitational constant.  

The above set of 3 mutually coupled, second-order, ordinary differential 
equations (ODEs) can be written explicitly in terms of the components of 
the vector $\vec R_i$, which means that there are 9 second-order ODEs; 
see \cite{Broucke1973} for an interesting and useful form of the equations.  
Since these equations can be converted into sets of 2 first-order ODEs, 
there are actually 18 first-order ODEs to fully describe the general 
three-body problem.  A standard mathematical procedure \citep{Whittaker1937} 
allows solving such ODEs by quadratures if independent integrals of 
motion exist.  Therefore, we shall now determine the number of integrals 
of motion for the general three-body problem \citep{McCord1998}. 

\subsection{Integrals of motion}
  
Let us sum Eq. (\ref{S2eq1}) over all three bodies and take into account 
the symmetry condition $\vec r_{ij} = - \vec r_{ji}$.  The result is
\begin{equation}
\sum_{i=1}^{3} M_i {{d^2 \vec R_i} \over {dt^2}} = 0\ ,
\label{S2eq2}
\end{equation}

\noindent
and after integration, we obtain
\begin{equation}
\sum_{i=1}^{3} M_i {{d \vec R_i} \over {dt}} = \vec C_1\ ,
\label{S2eq3}
\end{equation}

\noindent
where $\vec C_1$ = const.  One more integration yields
\begin{equation}
\sum_{i=1}^{3} M_i \vec R_i = \vec C_1 t + \vec C_2\ ,
\label{S2eq4}
\end{equation}

\noindent
with $\vec C_2$ = const.

Since the center of mass is defined as $\vec R_{cm} = \sum \limits_{i=1}^{3} 
M_i \vec R_i / \sum \limits_{i=1}^{3} M_i$, Eq. (\ref{S2eq4}) determines 
its motion, while Eq. (\ref{S2eq3}) shows that it moves with a constant 
velocity.  The vectors $\vec C_1$ and $\vec C_2$ are the integrals of 
motion, thus we have 6 integrals of motion by taking the components 
of these vectors.    

The conservation of angular momentum around the center of the inertial 
Cartesian coordinate system in the general three-body problem gives 
other integrals of motion.  To show this, we take a vector product of 
$\vec R_i$ with Eq. (\ref{S2eq2}), and obtain
\begin{equation}
\sum_{i=1}^{3} M_i \vec R_i \times {{d^2 \vec R_i} \over {dt^2}} = 0\ ,
\label{S2eq5}
\end{equation}

\noindent
which after integration yields
\begin{equation}
\sum_{i=1}^{3} M_i \vec R_i \times {{d \vec R_i} \over {dt}} = \vec C_3\ ,
\label{S2eq6}
\end{equation}

\noindent
where $\vec C_3$ = const.  Hence, there are 3 more integrals of motion.

An additional integral of motion is related to the conservation of the 
total energy of the system.  With the kinetic energy $E_{\rm kin}$ given 
by 
\begin{equation}
E_{\rm kin} = {1 \over 2} \sum_{i=1}^{3} M_i {{d \vec R_i} \over {dt}} 
\cdot {{d \vec R_i} \over {dt}}\ ,
\label{S2eq7}
\end{equation}

\noindent
and the potential energy $E_{\rm pot}$ defined as 
\begin{equation}
E_{\rm pot} = - {G \over 2} \sum_{i=1}^{3} \sum_{j=1}^{3} {{M_i M_j} 
\over {r_{ij}}}\ ,
\label{S2eq8}
\end{equation}

\noindent
where $i \neq j$, we have the total energy $E_{\rm tot} = E_{\rm kin} 
+ E_{\rm pot} \equiv C_4 = $const to be also an integral of motion. 

According to the above results, there are 10 (classical) integrals 
of motion in the general three-body problem, which means that the 
set of 18 first-order ODEs can be reduced to 8 first-order ODEs.  
Actually, there are 2 other integrals of motion, one related to the 
elimination of time and the other to the elimination of the so-called 
ascending node.  The time can be eliminated by transforming one of the 
dependent variables as an independent variable \citep{Szebehely1967,
BarrowGreen1997}.  \cite{Jacobi1843} changed variables so that two 
bodies would orbit a third one, and showed that the difference in 
longitude between the ascending nodes is fixed at $\pi$ radians,
which becomes an integral of motion \citep{BarrowGreen1997}.  
Having obtained the 12 integrals of motion, the system of 18 
equations can be reduced to 6 equations.  It has been proven 
that no other independent integrals of motion exist \citep
{Bruns1887,Poincare1892,Whittaker1937,Szebehely1967,Valtonen2006}. 

Since $E_{\rm kin} > 0$ and $E_{\rm pot} < 0$ (see Eqs (\ref{S2eq7})
(\ref{S2eq8}), respectively),  $E_{\rm tot}$ or $C_4$ can either be
positive, negative or zero, and $C_4$ can be used to classify motions
of the general three-body problem \citep{Roy2005}.  In the 
case $C_4 > 0$, the three-body system must split, which means that 
one body is ejected while the remaining two bodies form a binary 
system.  The special case of $C_4 = 0$ is unlikely to take place 
in Nature; however, if it indeed occurred, it would result in one 
body escaping the system.  Finally, the case $C_4 < 0$ may lead 
to either escape or periodic orbits with the result depending on 
the value of the moment of inertia given by $I = \sum \limits_{i=1}^3 
M_i R_i^2$; for details see \cite{Roy2005}, \cite{Marchal1990} and 
\cite{Valtonen2006}.   

In addition to the integrals of motions, the virial theorem 
$<E_{\rm kin}>$ = - $<E_{\rm pot}> / 2$, where $<E_{\rm kin}>$
and $<E_{\rm pot}>$ is the time average kinetic and potential
energy, respectively, can be used to determine the stability of 
the three-body problem and its statistical properties \citep
{Valtonen2006}.  The system is unstable if its time average 
kinetic energy is more than two times higher than its time 
average potential energy.  

\subsection{Another formulation}

The standard formulation of the general three-body problem 
presented in the previous section is often replaced by either the 
Hamiltonian formulation extensively used by \cite{Poincare1892}, 
or by the variational formulation described in detail and used 
by \cite{Siegel1991}.  Here, we present only the Hamiltonian 
formulation. 

Let us consider the so-called natural units and introduce $G = 1$ 
in Eqs (\ref{S2eq1}) and (\ref{S2eq8}).  Moreover, we write $\vec 
R_i = (R_{1i}, R_{2i}, R_{3i}) \equiv q_{ki}$, where $R_{1i}$, $R_{2i}$ 
and $R_{3i}$ are components of the vector $\vec R_i$ in the inertial
Cartesian coordinate system, and $k = 1$, $2$ and $3$.  Using this 
notation, we define the momentum, $p_{ki}$, as
\begin{equation}
p_{ki} = M_i {{d q_{ki}} \over {dt}}\ ,
\label{S2eq10}
\end{equation}

\noindent
and the kinetic energy as 
\begin{equation}
E_{\rm kin} = \sum_{k,i=1}^3 {{p_{ki}^2} \over {2 M_i}}\ ,
\label{S2eq11}
\end{equation}

\noindent
and introduce the Hamiltonian $H = E_{\rm kin} + E_{\rm pot}$ that 
allows us to write the equations of motion in the following Hamiltonian 
form
\begin{equation}
{{d q_{ki}} \over {dt}} = {{\partial H} \over {\partial p_{ki}}}
\hskip0.2in {\rm and} \hskip0.2in
{{d p_{ki}} \over {dt}} = - {{\partial H} \over {\partial q_{ki}}}\ ,
\label{S2eq12}
\end{equation}

\noindent
which is again the set of 18 first-order ODEs equivalent to the set of 
18 first-order ODEs given by Eq. (\ref{S2eq1}). 

The equations of motion describing the general three-body problem 
(either Eq. (\ref{S2eq1}) or Eq. (\ref{S2eq12})) can be further 
reduced by considering the general Hill problem, in which $M_1 >> 
M_2$ and $M_1 >> M_3$, however, $M_3$ is not negligible when 
compared to $M_2$.  The governing equations describing this 
problem are derived by \cite{Szebehely1967} and \cite{Simo2000}, 
who also discussed their applications to the Sun-Earth-Moon problem, 
known as the Hill lunar theory.  In this paper, we limit our discussion 
of the Hill problem to its circular and elliptic versions of the restricted
three-body problem, and describe them in Section 6.5 and 7.3, respectively. 

\subsection{General properties of solutions of ODEs}

Solutions of the ODEs describing the general three-body problem can 
be represented by curves in 3D.  In a special case, a solution can be 
a single point known also as a {\it fixed point} or as an equilibrium 
solution; a trajectory can either reach the fixed point or approach 
it asymptotically.  Typically, periodic orbits are centered around 
the fixed points, which are called stable points or stable centers.  
Moreover, if a trajectory spirals toward a fixed point or moves away 
from it, the point is called a spiral {\it sink} or spiral {\it source}, respectively.  There can also be a {\it saddle point} in which two 
trajectories approach the point and two leave it, but all other 
trajectories are kept away from it.    

From a mathematical point of view, it is required that solutions to 
ODEs exist and that they are unique.  The existence can be either 
global when a solution is defined for any time in the past, present 
and future, or local when it is defined only for a short period of 
time.  The uniqueness of solution means that there is only one 
solution at each point.  For given ODEs, the existence and uniqueness 
are typically stated by mathematical theorems.  Let us consider the
following first-order ODE $y^{\prime} (x) = f (y(x),x)$, where 
$y^{\prime} = dy / dx$, and the intial condition is $y (x_0) = y_0$, 
with $x_0$ being the initial value of $x \epsilon [x_0 - \varepsilon,
x_0 + \varepsilon]$.  Picard's (or Lipschitz and Cauchy's) existence
theorem states that if $f (y(x),x)$ is a Lipschitz continuous function
in $y$ and $x$, then there is $\varepsilon > 0$ such that a unique 
solution $y(x)$ exists on the interval $[x_0 - \varepsilon,x_0 + 
\varepsilon]$.
   
Another requirement is that the problem is well-defined, which means
that a solution must be continuous with respect to the initial data.
Being aware of the above mathematical requirements, Poincar\'e studied 
the problem of solving ODEs from another perspective by developing 
qualitative methods, which he described in the third volume of his 
book \citep{Poincare1892};  Poincar\'e's work was done mainly for 
the CR3BP, and we describe it in Section 6.

\section{Analytical studies of the general three-body problem}

\subsection{Euler and Lagrange periodic solutions}

As already mentioned in Section 1, two different classes of periodic 
solutions to the general three-body problem were obtained by 
\cite{Euler1767} and \cite{Lagrange1772} who considered a linear 
and triangle configuration of the three bodies, respectively, and 
demonstrated that in both cases the bodies move along elliptic orbits.  
Since the masses of the three bodies are arbitrary (finite), the problem 
is considered here as the general three-body problem despite the 
limitation on the bodies mutual positions; because of this limitation, 
the Euler and Lagrange solutions are called the particular solutions.  
The existence of the special configurations (known as central 
configurations in celestial mechanics) plays an important role in 
studies of orbital stability. Detailed derivations of the Euler and 
Lagrange solutions are given by \cite{Danby1988}. 

\begin{figure}
\includegraphics[width=\linewidth]{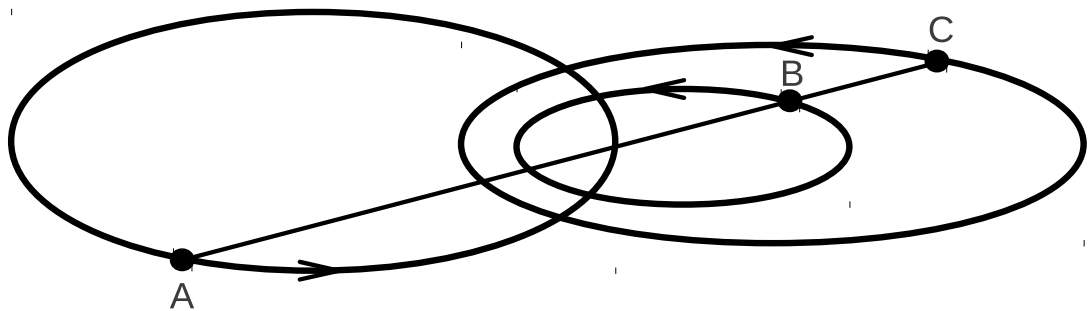}
\caption{Illustration of the Euler solution with a line joining three 
masses $M_1$, $M_2$ and $M_3$ located at the corresponding points A, 
B, and C.}
\label{fig:Euler}
\end{figure}
      
In the Euler solution, the initial linear configuration of the three 
bodies is maintained (it becomes the central configuration), if the 
ratio AB/BC (see Fig. \ref{fig:Euler}) has a certain value that depends 
on the masses, and if suitable initial conditions are specified.  Euler 
proved that the line AC would rotate about the center of mass of the 
bodies leading to periodic motions of all three bodies along ellipses.  
Moreover, he also demonstrated that the ratio AB/BC would remain 
unchanged along AC during the motion of the bodies.  Since the three 
bodies can be ordered in three different ways along the line, there 
are three solutions corresponding to the ordering of the bodies.  
However, it must be noted that the Euler solution is unstable against 
small displacements.

\begin{figure}
\includegraphics[width=\linewidth]{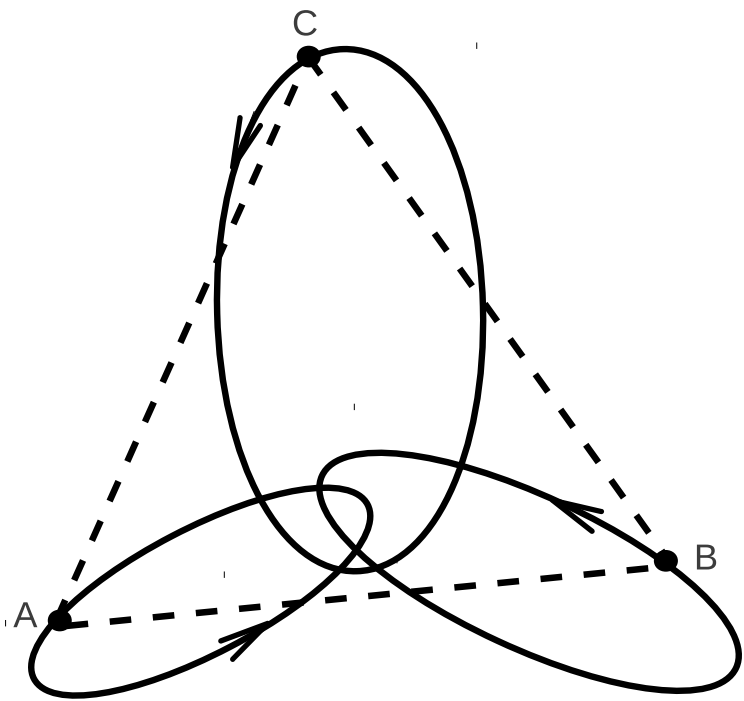}
\caption{Illustration of the Lagrange solution with an equilateral 
triangle joining three masses $M_1$, $M_2$ and $M_3$ located at the 
corresponding points A, B, and C.}
\label{fig:Lagrange}
\end{figure}

Now, in the Lagrange solution, the initial configuration is an 
equilateral triangle and the three bodies are located at its vertices.  
Lagrange proved that for suitable initial conditions, the initial 
configuration is maintained (becomes the central configuration) and 
that the orbits of the three bodies remain elliptical for the duration 
of the motion (see Fig. \ref{fig:Lagrange}).  Despite the fact that the 
central configuration is preserved, and the triangle changes its size 
and orientation as the bodies move, the triangle remains equilateral.  
Since the triangle can be oriented in two different ways, there are 
two solutions corresponding to the orientation of the triangle.  
Regions of stability and instability of the Lagrange solutions 
were identified by \cite{Mansilla2006}.

\subsection{Other periodic solutions}

\cite{Hill1877,Hill1878} found periodic orbits in the problem that is now 
known as the Hill problem (see Section 1).  \cite{Poincare1892} developed 
a method that allowed him to find periodic orbits in the CR3BP (see Section
6).  However, the original version of his method can also be applied to the 
general three-body problem.  We follow \cite{Poincare1892} and 
consider the autonomous Hamiltonian system given by 
\begin{equation}
{{d x_i} \over {dt}} = {{\partial F} \over {\partial y_i}}
\hskip0.5in {\rm and} \hskip0.5in
{{d y_i} \over {dt}} = - {{\partial F} \over {\partial x_i}}\ ,
\label{S3eq4}
\end{equation}

\noindent
and $F = F_0 + \mu F_1 + \mu F_2 + ...$, where $F_0 = F_0 (x)$, however, 
$F_1$, $F_2$, ... are functions of both $x$ and $y$ and they are periodic 
(with a period of $2\pi$) with respect to $y$.  Moreover, $\mu$ is a parameter 
that depends on mass, where $i=1$, $2$, and $3$.  If $\mu = 0$, then $x_i$ = 
constant and $y_i = \eta_i t + \omega_i$, with $\eta_i = \partial F_0 / 
\partial x_i$ and $\omega_i$ being constants of integration, and a solution 
is periodic when the values of $\eta_i$ are commensurable;  actually, if 
$\mu = 0$, then there are infinitely many constants $x_i$ that lead to 
periodic solutions \citep{BarrowGreen1997}.  

Then Poincar\'e wanted to know whether the periodic solutions could be 
analytically continued when $\mu$ remains small.  The approach presented 
below follows \cite{Poincare1890}, in which he assumed that the $\mu 
\neq 0$ solutions at $t=0$ are:  $x_i = a_i + \delta a_i$ and $y_i = 
\omega_i + \delta \omega_i$, with $a_i$ being constants.  Now, for $t=T$, 
where $T$ is the lowest common multiple of the $2\pi / \eta_i$, the 
solutions are: $x_i = a_i + \delta a_i + \Delta a_i$ and $y_i = \omega_i 
+ \eta_i T + \delta \omega_i + \Delta \omega_i$, which means that they 
are periodic if all $\Delta a_i = 0$ and all $\Delta \omega_i = 0$.  
Since only five equations given by Eq. (\ref{S3eq4}) are independent 
(as $F$ = const is their integral), Poincar\'e showed that the five 
equations could be satisfied if   
\begin{equation}
{{\partial \Psi} \over {\partial \omega_2}} = {{\partial \Psi} \over 
{\partial \omega_3}} = 0\ ,
\label{S3eq5}
\end{equation}

\noindent
where $\Psi$ is a periodic function with respect to $\omega_1$ and
$\omega_2$, and if the following additional conditions are also 
satisfied: the Hessian of $\Psi$ with respect to $\omega_1$ and
$\omega_2$ must be non-zero and the Hessian of $F_0$ with respect 
to $x_i^0$ must also be non-zero; the so-called Hessian condition
is ${\rm Det}(\partial^2 \Psi / \partial \omega_1 \partial \omega_2) 
\neq 0$ and similar for $F_0$.  Moreover, if $\eta^{\prime}_i = \eta_i 
(1 + \varepsilon)$, where $\varepsilon$ is small, then there exists 
the following periodic solutions $x_i = \phi_i (t,\mu,\varepsilon)$ 
and $y_i = \phi^{\prime}_i (t,\mu,\varepsilon)$ with period $T^{\prime} 
= T / (1 + \varepsilon)$.

Using this method, Poincar\'e established that there were periodic 
orbits for all sufficiently small values of $\mu$.  Poincar\'e's 
approach was generalized first by the Poincar\'e-Bendixson theorem
\citep{Bendixson1901}, and by the Poincar\'e-Birkhoff theorem 
\citep{Birkhoff1912,Birkhoff1913,Birkhoff1915}.  According to these 
theorems, there are periodic solutions even for non-integrable 
Hamiltonian systems.  However, the theorems do not address a problem 
of what happens to solutions of an integrable Hamiltonian system when 
the governing equations are slightly perturbed?  The answer to this 
question is given by the KAM theorem \citep{Kolmogorov1954,Moser1962,
Arnold1963} described in Section 3.3.

An important result concerning the existence of periodic solutions 
in the general three-body problem was obtained by \cite
{Hadjidemetriou1975a}, who proved that any symmetric periodic orbit 
of the CR3BP could be continued analytically to a periodic orbit in 
the planar general three-body problem \citep{Hadjidemetriou1975b}.  
Using this analytical result, families of periodic orbits in the 
planar general three-body problem were constructed \citep{Bozis1976} 
and their stability was investigated \citep{Hadjidemetriou1975c}.  
Moreover, \cite{Katopodis1979} demonstrated that it was possible to 
generalize Hadjidemetriou's result from the 3D CR3BP to the 3D general 
three-body problem.  

\begin{figure}
\includegraphics[width=\linewidth]{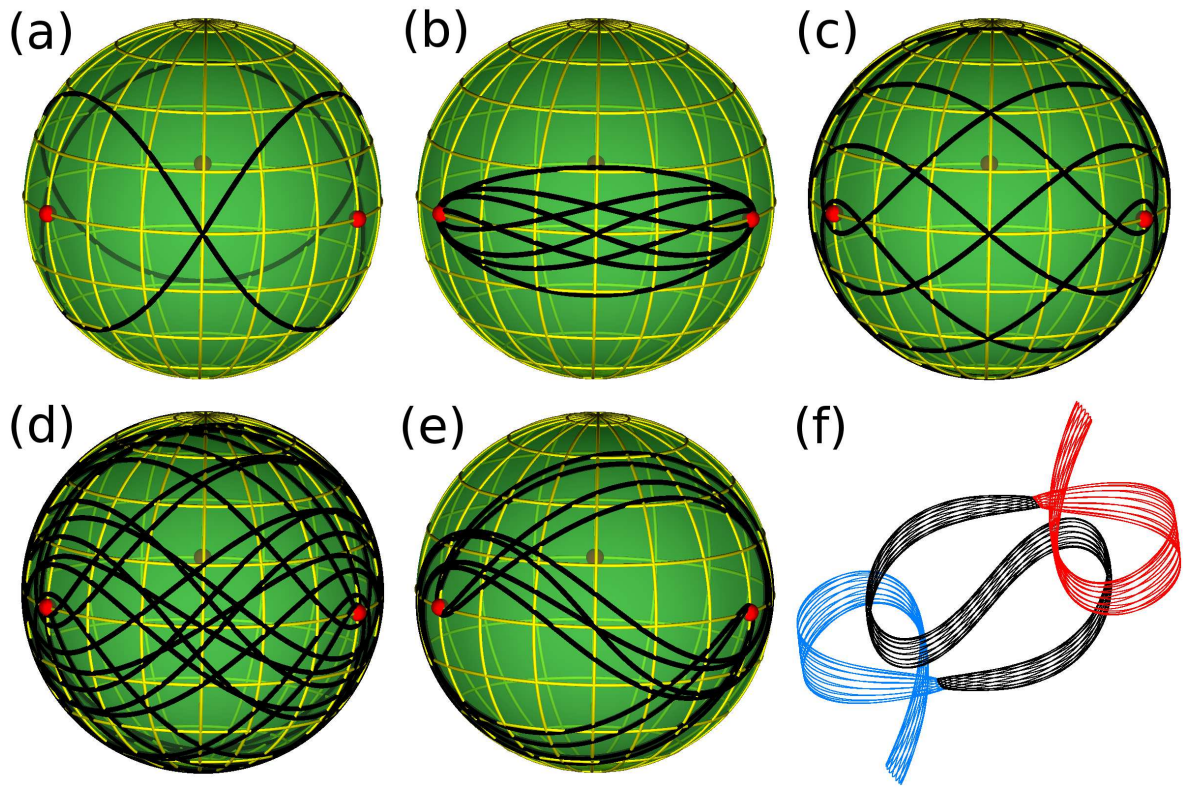}
\caption{New periodic solutions are presented on the shape-space sphere 
with its back sides also visible. (a) The figure-8 orbit; (b) Butterfly 
orbit; (c) Moth orbit; (d) Yarn orbit; (e) Yin-yang orbit; (f) The 
yin-yang orbit in real space.  Reproduced after \cite{Suvakov2013a}.
Copyright 2013 Physical Review Letters.}
\label{fig:Suvakov}
\end{figure}

Periodic orbits in the 3D general three-body problem are typically 
determined numerically \citep{Henon1965,Henon1974,Szebehely1967,
Moore1993,Chenciner2000,Suvakov2013a}.  To make the computations as 
efficient as possible, the 3D general three-body problem was formulated 
suitably for numerical computations \citep{Markellos1980} by using an 
idea of \cite{Hadjidemetriou1975c}.  Periodic solutions with binary 
collisions in the general three-body problem were also determined 
analytically for different masses, and 8 different periodic orbits 
in a rotating frame of reference were found by \cite{Delibaltas1983}.  
By starting with periodic orbits for the 3D CR3BP, families of periodic 
orbits in the 3D general three-body problem were numerically determined, 
and it was shown that the latter was not isolated but instead they form 
continuous mono-parametric families for given masses of the three bodies 
\citep{Markellos1981}.  Let us also point out that there are methods for constructing analytically periodic solutions to the general three-body 
problem by using computer experiments \citep{Valtonen1989}.

More recently, new periodic solutions to the three-body problem were found 
using combinations of analytical and numerical methods.  Specifically, 
\cite{Moore1993} discovered, and \cite{Chenciner2000} independently 
re-discovered, the so-called 8-type periodic solution for the general 
three-body problem with equal masses (see Fig. \ref{fig:Suvakov}a), 
and formally proved its existence \citep{Montgomery2001}.  A search 
for periodic orbits in the vicinity of the 8-type periodic solution 
was performed numerically by \cite{Suvakov2013b}.  The KAM theorem 
was used by \cite{Simo2002} to establish stability of this solution, 
and a rotating frame of reference was adapted to study properties of 
the solution \citep{Broucke2006}.  Moreover, new symmetric non-collision 
periodic solutions with some fixed winding numbers and masses were 
found by \cite{Zhang2004}. 

Currently, 13 new distinct periodic orbits were found by \cite{Suvakov2013a}, 
who considered the planar general three-body problem with equal masses and 
zero angular momentum (see Fig. \ref{fig:Suvakov}b,c,d,e,f).  The authors
presented a new classification of periodic solutions but admitted that 
their results could not be verified observationally because so far no 
astronomical systems with the considered properties of the three bodies 
are known.     

\subsection{The KAM theorem and stability of periodic solutions}

Let us write Eq. (\ref{S2eq12}) in the following general form
\begin{equation}
{{d q} \over {dt}} = {{\partial H} \over {\partial p}}
\hskip0.2in {\rm and} \hskip0.2in
{{d p} \over {dt}} = - {{\partial H} \over {\partial q}}\ ,
\label{S3eq6}
\end{equation}

\noindent
where $q = q_1$, ..., $q_n$, $p = p_1$, ..., $p_n$, where $t$ is time. 
It means that the equations describe the autonomous Hamiltonian system 
with $n$ degrees of freedom.  The Hamiltonian is $H (p,q) = H_0 (p) + 
\mu H_1 (p,q) + ...$, with $H$ being periodic in $q$ with period $2\pi$, 
and $\mu$ is a small parameter.  

For the unperturbed motion $\mu = 0$, the above equations reduce to  
\begin{equation}
{{d q} \over {dt}} = {{\partial H_0} \over {\partial p}} = \Omega(p)
\hskip0.2in {\rm and} \hskip0.2in
{{d p} \over {dt}} = 0\ ,
\label{S3eq7}
\end{equation}

\noindent
with $\Omega = \Omega_1$, ..., $\Omega_n$.  Clearly, Eq. (\ref{S3eq4}) 
can be integrated where the resulting trajectories are confined to a 
torus in the phase space, actually, there are invariant tori $p$ = 
const; see \cite{BarrowGreen1997} and \cite{Diacu1996} for details.
For incommensurable frequencies the motion is quasi-periodic.  Moreover, 
the system is nondegenerate because the Hessian determinant is not zero.

Now, the KAM theorem tells us that when the system is slightly perturbed
most of the invariant tori are not destroyed but only slightly shifted in
the phase space.  This has important implications on stability of orbits 
in the general and restricted three-body problem.  The proof of the KAM 
theorem by \cite{Moser1962} and \cite{Arnold1963} also demonstrated that 
convergent power series solutions exist for the three-body (as well as 
for the n-body) problem.  The KAM theorem seems to be very useful for 
studying the global stability in the three-body problem \citep
{Robutel1993a,Montgomery2001,Simo2002}; however, some of its applications 
are limited only to small masses of the third body and as a result, 
different methods have been developed to deal with the general 
three-body problem \citep{Robutel1993b}. 

\subsection{Collision singularities and regularization}

The Euler, Lagrange and other periodic solutions require suitable initial 
conditions, so an interesting question to ask is:  What happens to the 
three bodies when the initial velocities are set equal to zero?  Apparently, 
in this case all three bodies move toward their center of mass and undergo 
a triple collision at that point in finite time.  At the point where the 
triple collision occurs, the solutions (if they are known) become abruptly 
terminated, which means that we have a triple collision singularity at that 
point.  Similarly, we may have a double collision singularity when two 
bodies collide.  In the following, we describe collision singularities 
in the general three-body problem and present a method called 
regularization to remove them. 

Based on the important work by \cite{Painleve1896,Painleve1897}, it was 
established that collisions between either two or three bodies are the only singularities in the three-body problem, and that such singularities could 
be removed by setting certain sets of initial conditions.  The structure 
of phase space near the singularities changes and solutions (if they exist) 
do not end at the singularities but come close to them and show strange behaviours.  The work of Painlev\'e on singularities resulting from 
collisions between two bodies was continued by \cite{LeviCivita1903} and 
\cite{Bisconcini1906}, with the former concentrated on the restricted 
three-body problem and the latter on the general three-body problem.  
They addressed an important problem of initial conditions that lead to 
collisions, and discussed regularization that allows extending possible 
solutions beyond a singularity.  Bisconcini deduced two analytical 
relationships between initial conditions and proved that when the 
relationships were satisfied a collision did take place in a finite 
time.

The problem of triple collisions was investigated by \cite{Sundman1907,
Sundman1909}, who formulated and proved two important theorems.  First 
he proved that the triple collisions would occur only if all the constants 
of angular momentum are zero at the same time.  Second, he demonstrated 
that if all three bodies collide at one point in space they move in the 
plane of their common center of gravity, and as they approach the collision 
they also approach asymptotically one of the central configurations, namely, 
either Euler's collinear configuration or Lagrange's equilateral triangle.  
Proofs of Sundman's theorems and their applications are described in detail 
by \cite{Siegel1991}.  \cite{Sundman1912} also studied binary collisions 
and used the results of those studies to find a complete solution to the 
general three-body problem (see Section 3.3).  Moreover, 
\cite{Waldvogel1980} considered the variational equations of the three-body 
problem with triple collisions and obtained solutions given in terms of hypergeometric functions, which are valid in the vicinity of the collision. 
   
Regularization means that the motion is extended beyond a singularity, 
which occurs in a solution, through an elastic bounce and without any loss 
or gain of energy.  Thus, it is important to know whether such a solution 
with a singularity can always be continued in a meaningful way?  
\cite{Siegel1941} showed that in general this cannot be done by proving a 
theorem, which states that an analytical solution that goes through the 
triple collision cannot be found for practically all masses of the bodies.
The work of Siegel was significantly extended by \cite{McGehee1974}, who 
obtained new transformations, now known as McGehee's transformations.  He 
also introduced a new coordinate system that allowed him to magnify the 
triple collision singularity and inspect it more closely.  What he found 
was a distorted sphere with four horns extending to infinity and he named 
this surface a collision manifold; for more details see \cite{Diacu1996}.  
There are two currently used regularization schemes, the so-called K-S 
regularization scheme developed by \cite{Kustaanheimo1965}, and the B-H regularization scheme developed by \cite{Burdet1967} and \cite{Heggie1973,
Heggie1976}.  These two analytical schemes describe close two-body encounters 
and they are typically used to supplement numerical simulations 
\citep{Valtonen2006}.

\begin{figure}
\centering
\begin{tabular}{cc}
\includegraphics[width=0.5\linewidth]{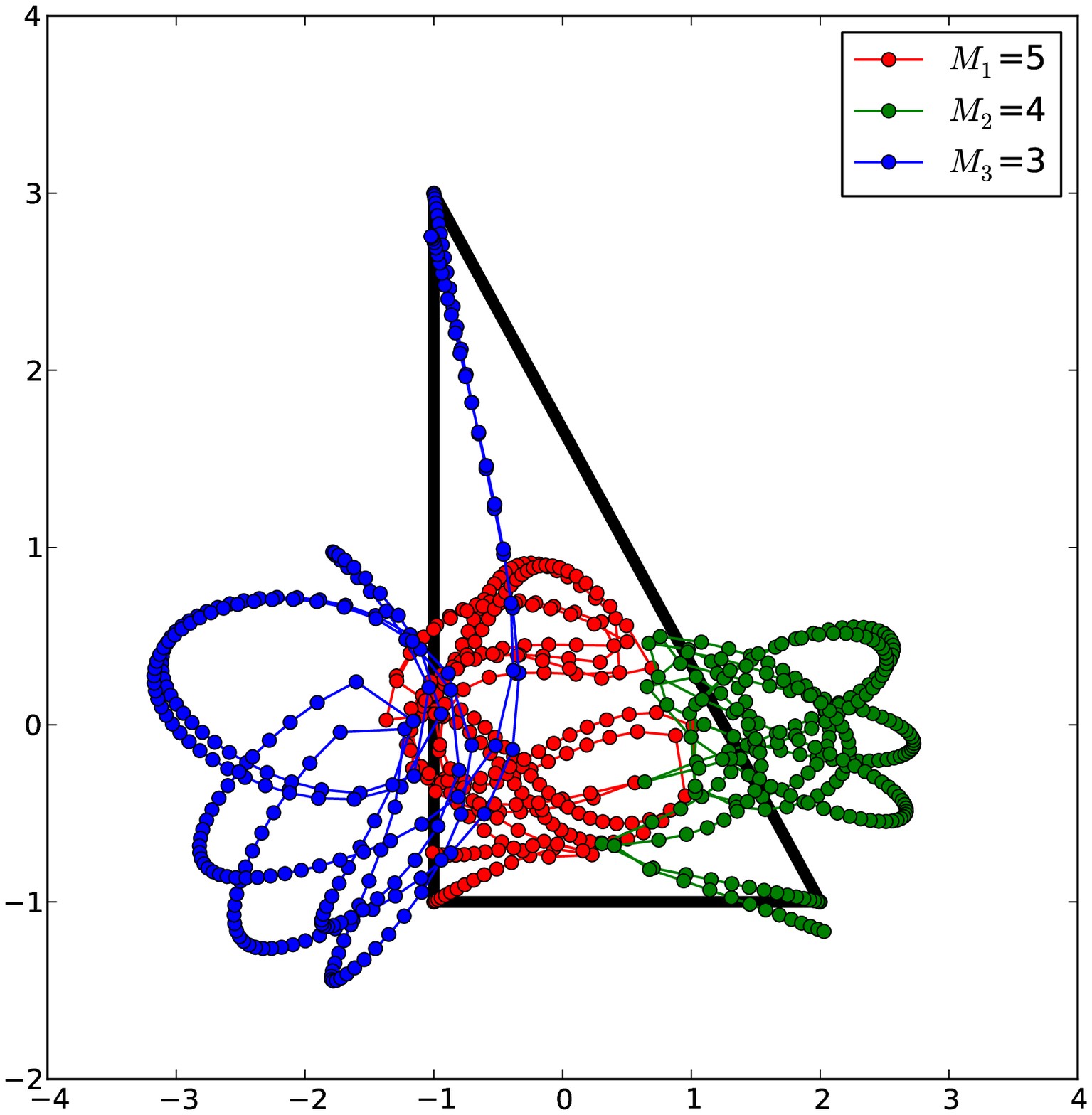}&
\includegraphics[width=0.5\linewidth]{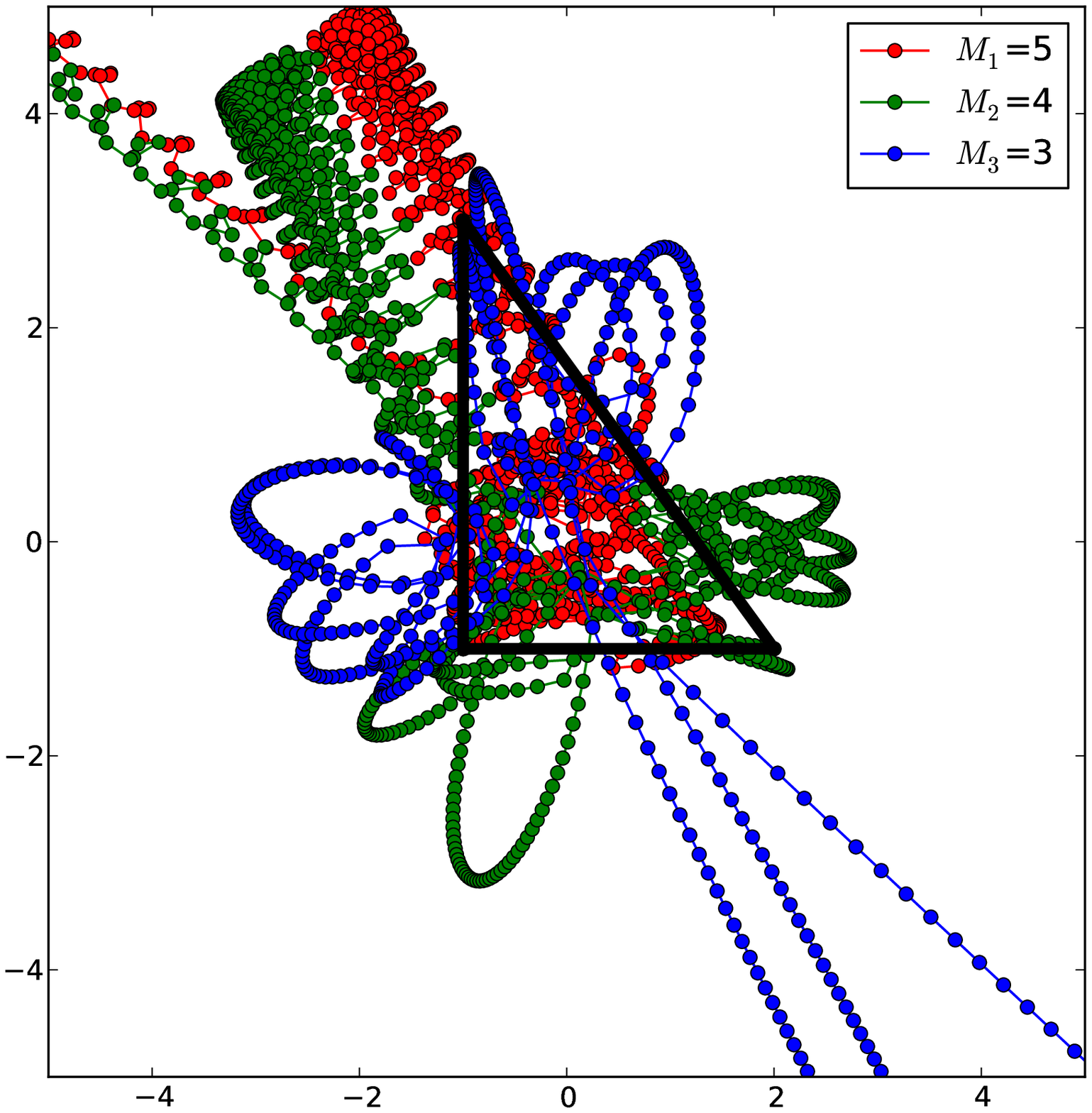}
\end{tabular}
\caption{Numerical solutions to the Pythagorean triangle problem 
show the evolution of the system for the first 10 orbits (left) 
and up to the ejection of the smallest mass from the system (right).}
\label{fig:Szebehely}
\end{figure}

In Section 1, we briefly discussed the numerical work by \cite{Szebehely1967a,Szebehely1967b}, who considered three bodies with their masses proportional 
to 3, 4 and 5 located at the vertices of a Pythagorean triangle with the sides 
equal to 3, 4 and 5 length units, and showed that two of these objects form a 
binary system but the third one is expelled from the system (see Fig. 4).  
Such a pure numerical result required an analytical verification and a formal mathematical proof that was given by \cite{McGehee1974} for the three-body problem 
in which the bodies are restricted to move only along a fixed line.  A more 
general proof for the planar three-body problem was given by \cite{Waldvogel1976}.  The main result of both proofs was that after a close triple encounter one of 
the bodies of the three-body problem would escape from the system.  Moreover, sufficient conditions for the escape of one body from the general three-body 
system can also be specified \citep{Standish1971,Anosova1994} as well as the conditions that an ejected body would return to the system \citep{Standish1972}. 

\subsection{Sundman's complete solutions}

\cite{Painleve1896,Painleve1897} formulated a conjecture that once collisions 
are excluded by choosing certain initial conditions, then the equations of 
motion of the general three-body problem could be integrable using power 
series solutions.  Painlev\'e neither proved his conjecture nor found such solutions.  Nevertheless, his results had strong impact on his contemporaries 
who searched extensively for the solutions but failed to find them.  The only
one who succeded was \cite{Sundman1907} who found a complete solution to the
general three-body problem.

As already stated in Section 1, searches for power series solutions had 
been pursued before Painlev\'e formulated his conjecture.  Different
power series solutions were presented and their convergence rates were
investigated \citep{Delaunay1867,Lindstedt1884,Gylden1893}.  Some important 
points related to the convergence of those series were made \citep{Poincare1892} 
but he also did not succeed in finding the solution.  A breakthrough occurred 
when \cite{Sundman1907,Sundman1909,Sundman1912} studied double collision singularities.  Let us now briefly describe Sundman's complete solutions to 
the general three-body problem by following \cite{BarrowGreen1997}.

Sundman considered a double collision singularity and showed that it
could be removed by introducing a regularizing variable $u$ defined
in terms of time $t$ as
\begin{equation}
u = \int_{t_0}^t {{dt} \over {t}}\ ,
\label{S3eq1}
\end{equation}

\noindent
where $t_0$ corresponds to $u=0$ for which the system is regular.  With 
the singularity occurring at $t=t_1$, Sundman discovered that the coordinates 
of the three bodies could be expanded in powers of $(t-t_1)^{1/2}$ and that 
he could use an analytical continuation to demonstrate that the expansion 
described correctly motions after the collision.  The energy remained 
unchanged, the areal velocity stayed constant, and the solutions were 
valid for $t > t_1$ and described motions correctly after each new double 
collision; note that the triple collisions were not allowed in this approach.

Since Sundman's regularization transformation was dependent on $t_0$, he 
introduced another variable, namely, $dt = \Gamma d\omega$, where 
$\Gamma = (1 - e^{-r_0/l})(1 - e^{-r_1/l})(1 - e^{-r_2/l})$ and $r_0$,
$r_1$ and $r_2$ are the mutual distances, with the two greater of them
being larger than $l$, which gives $0 \leq \Gamma \leq 1$, so there 
is a one-to-one correspondence between $t$ and $\omega$. 

If $\omega^*$ is a real and finite value of $\omega$, then the coordinates
of the three bodies, their mutual distances and time can be expanded in 
power series with respect to $(\omega - \omega^*)$, with the radius of 
convergence $\Omega$ satisfying $\vert \omega - \omega^* \vert \leq \Omega$.
Finally, Sundman defined  
\begin{equation}
\tau = {{e^{\pi \omega / 2 \Omega} + 1} \over {e^{\pi \omega / 2 \Omega} + 1}}\ ,
\label{S3eq2}
\end{equation}
and used
\begin{equation}
\omega = {{2 \Omega} \over {\pi}} {\rm log} {{1 + \tau} \over {1 - \tau}}\ ,
\label{S3eq3}
\end{equation}

\noindent
which allowed him to transform the $\omega-$plane into a circle of unit radius
in the $\tau-$plane.  This guarantees that the coordinates of the bodies, their 
mutual distances and time are analytic functions of $\tau$ everywhere inside the
circle.  Moreover, the expansions are convergent and their different terms can 
be calculated once $l$ and $\Omega$ are specified.  Thus, these are complete 
solutions to the general three-body problem.

Clearly, this is a remarkable result and this fact had been recognized by 
Sundman's contemporaries \citep{LeviCivita1918,Birkhoff1920}.  However, 
there is a major problem with Sundman's solutions; namely, their convergence 
is extremely slow, actually so slow that it requires millions of terms to 
find the motion of one body for insignificantly short durations of time.  
Since the solutions are also not useful for numerical computations, owing 
to the round-off errors, they have no practical applications.  Thus, we 
have a paradox: though we know the complete solutions, nothing is added 
to the previously accumulated knowledge about the three-body problem!  
A more detailed discussion of Sundman's complete solutions and his other 
relevant results can be found in \cite{Siegel1991} and \cite{BarrowGreen1997}. 

\subsection{Zero velocity hypersurfaces}

As already mentioned in Section 1, \cite{Hill1877,Hill1878} introduced 
the zero velocity curves (ZVC) for the restricted three-body problem; 
these curves are boundaries of regions in space where the bodies are 
allowed to move.  Hill's results can be generalized to the 3D general 
three-body problem; however, in this case, regions of space where 
motions are disallowed are not curves but rather 4D surfaces, or 
hypersurfaces.  The so-called zero velocity hypersurfaces (ZVH) can 
be determined using Sundman's inequality \citep{Sergysels1986} and 
the resulting restrictions on the motions must obviously be consistent 
with the existing solutions to the problem \citep{Marchal1990,Valtonen2006}.  
If the three bodies have masses $M_1 > M_2 > M_3$, and if the terms of 
the order $(M_3 / M_2)^2$ can be neglected, then the ZVH of the general 
three-body problem become the ZVC of the restricted three-body problem 
\citep{Milani1983}.

\section{Numerical solutions to the general three-body problem}

Since the analytical solutions to the three-body problem obtained by 
\cite{Sundman1907} converge extremely slowly, the modern practice of 
solving this problem involves numerical computations.  In general, 
there are different numerical methods to solve the three-body problem,
and determine periodic orbits, resonances and chaotic indicators.  
In the following, we present an overview of these methods, including 
numerical simulation packages such as \textit{MERCURY}, \textit{SWIFT}, 
and \textit{HNBody} developed for astronomical applications \citep{Chambers1999,Levison1994,Rauch2002}.

\subsection{Numerical methods}

One of the commonly used methods in determining numerical solutions is 
the application of symplectic integration.  This method relies on the 
mathematical parametrization of the equations of motion into matrices 
and assumes a low occurrence of events where the orbital velocity could 
quickly become large.  In celestial mechanics, this is an implicit 
assumption on the likely values of eccentricity for the system being 
considered.  With low eccentricity, larger time steps can be implemented, 
which can significantly quicken the calculations.  This is possible due to 
the symplectic nature of Hamiltonian dynamics where matrices are symmetric 
and easily invertible.  Using this mathematical property, one can study 
nonlinear Hamiltonian dynamics with computationally fast mapping integration methods.  \cite{Wisdom1992} developed an algorithm that incorporates this 
idea and applied it to study the n-body problem.  

In general, one can envision a basic mapping method such as Euler's or 
leapfrog's integration, implemented with symplectic properties to attain 
higher speed and accuracy over more computationally expensive methods 
such as the Runge-Kutta method \citep{Press1992}.  We now briefly outline 
the leapfrog method, and then demonstrate how it can be transformed into 
a symplectic method.

The leapfrog integration method applies to problems that can be expressed 
as a second order differential equation of the following form:
\begin{align}
\ddot{x} &= F(x).
\end{align}
Many celestial mechanics problems are of this form and could thus be easily 
solved using the leapfrog integration method, which is stable for oscillatory 
motion as long as the time step, $\Delta t$, is constant and less than $2/
\omega$; this parameter is related to the mean motion $\omega = \sqrt{{G(M+m) 
/ a^3}}$, where $a$ is a semi-major axis.  For example, the numerical 
integration of Solar System dynamics has a maximum time step on the order 
of 0.0767 years.  However, in practice, one will use a smaller time step 
to better sample Mercury's orbit.  The choice of $\Delta t$ will be at 
least an order of magnitude larger than other numerical methods and hence computationally quicker.  The leapfrog method of integration utilizes half 
step evaluations of the following form:
\begin{align}
x_i &= x_{i-1} + v_{i-1/2}\Delta t, \nonumber \\
a_i &= F(x_i), \nonumber \\
v_{i+1/2} &= v_{i-1/2} + a_i\Delta t,
\end{align}
where $a_i$ are coefficients determined by $F(x_i)$, and $v_i$ is 
the velocity.

The main advantage of this method is its symplectic properties, which 
inherently conserve the orbital energy of the system.  This is highly 
valued in the field of celestial mechanics.  Also the leapfrog method 
is time-reversible, which means that one can integrate forward $n$ 
steps to a future state and reverse the direction of integration to 
return to the initial state.

Other commonly used computational techniques can be categorized broadly 
as Runge-Kutta like methods.  Different approaches to the general algorithm 
of Runge-Kutta have produced methods of determining the coefficients for 
the Runge-Kutta method to high order but with a decrease in computational 
speed.  Such approaches include Bulirsch-Stoer, Dormand-Prince, and 
St\"ormer-Cowell techniques \citep{Press1992}, where variable time steps 
along with sophisticated predictor-correctors have been utilized to make 
them more adaptive and accurate when compared with symplectic methods.  
We show the basic approach of the Runge-Kutta-Fehlberg (RKF) implementation 
following \cite{Press1992}.  The basic generalization is
\begin{align}
y_{n+1} &= y_n + \sum_{i=1}^s b_i\;k_i, \nonumber \\
{\rm where}& \nonumber \\
k_1 &= h\;f(t_n,y_n) \nonumber \\
k_2 &= h\;f(t_n+c_2h,y_n+a_{21}k_1) \nonumber \\
&\dots \nonumber \\
k_s &= h\;f(t_n + c_sh, y_n + a_{s1}k_1 + a_{s2}k_2 + \dots + 
a_{s,s-1}k_{s-1}),
\end{align}
where $y_n$ is a dependent variable, $t_n$ is an independent variable, 
$s$ is a number of stages, and the coefficients $b_i$, $c_i$ and $a_{ij}$, 
with $i = 1,2,...,s$ and $1 \leq j < i \leq s$, are determined from the 
so-called Butcher tableau, which give the relationships between these 
coefficients.  The most common RKF implementation is RKF45, which 
describes a method with $s=6$.  The adaptive portion of the method 
compares an order $m$ method ($y_{n+1}$) with an order $m-1$ approach
($y_{n+1}^*$), where $m=5$ for RKF45.  This comparison is used to 
determine the relative error between the orders, which will be a 
criterion to modify the step-size $h$.  The error estimate $e_{n+1}$ 
with the control scalar $z$ is of the following form
\begin{align}
e_{n+1} &= y_{n+1} - y_{n+1}^* = \sum_{i=1}^s (b_i - b_i^*)k_i, \nonumber \\
z &= \left({\epsilon\:h \over 2 \left|e_{n+1}\right|}\right)^{1\over 4}.
\end{align}

In order to determine the value of $z$, one specifies the error control 
tolerance $\epsilon$.  Then, the product of $z$ with the step-size $h$ is
used as the new optimal step-size $h^\prime$ for the next iteration.  When considering simulations where close encounters are highly probable, an 
adaptive step-size method is preferred.  Otherwise large errors in the 
total energy can accrue and impact the reliability of the results.

A more classical method of integration uses the Taylor method, which 
relies on the knowledge of terms within a Taylor series.  In general, 
the optimal number of terms in the Taylor series may not be known, thus optimization packages such as {\it TAYLOR} uses a text input of the 
ODEs to be solved and provides the user with an optimal time-stepper 
that will minimize error \citep{Jorba2005}.  This method can be 
comparable in terms of speed and accuracy with the other methods 
presented; however implicit assumptions are made that would need 
to be evaluated before a specific application.

\subsection{Numerical search for periodic orbits and resonances}

Periodic orbits have been determined analytically for special cases with the 
Euler and Lagrange solutions being the most notable.  However, other analytical solutions have also been found in other periodic families by using numerical 
methods (see \cite{Suvakov2013b} for details).  Studies of periodic orbits are 
related to the topic of resonance.  \cite{Mardling2008} and \cite{Murray1999} 
give detailed descriptions of the perturbation theory involved in determining 
the locations of resonance based on approximations to a harmonic oscillator.  
Using the perturbation theory, appropriate initial guesses can be made as to 
the locations of periodic orbits and resonances within a parameter space of 
semi-major axis and eccentricity.  Such guesses are made using the ratio of 
Kepler's harmonic law between two orbiting bodies.  

The nominal resonance location, $R_a$, can be defined in terms of two integers 
($k,l$), where $l$ is the order the resonance to form a ratio of $k / (k+l)$.  
For internal resonances, the third body orbits between the primary and 
secondary, and the relation is given by
\begin{equation}
R_a = \left(k \over k + l\right)^{2/3}\:\left(m_p \over m_p + 
m_s\right)\:a
\label{S4eq1}
\end{equation}
where $a$, $m_p$, and $m_s$ denote the semi-major axis of the secondary, 
mass of the primary, and mass of the secondary, respectively.  There is 
a similar definition for the external resonances.

Knowing this relation, one can begin to search for resonances that lead 
to periodic orbits using the general form of the resonant argument as
\begin{equation}
\phi = j_1\lambda^{\prime} + j_2\lambda + j_3\varpi^{\prime} + j_4\varpi 
+ j_5\Omega^{\prime} + j_6\Omega\ ,
\label{S4eq2}
\end{equation}
with conditions on the $j_n$ integer coefficients using the d'Almebert rules 
\citep{Hamilton1994}, which require a zero sum of the coefficients and an 
even sum for the $j_5$ and $j_6$ coefficients due to symmetry conditions.  
Also, the $\lambda^{\prime}$, $\varpi^{\prime}$, $\Omega^{\prime}$ represent 
the mean longitude, longitude of pericenter, and ascending node of the perturber.  Accounting for all these conditions, a numerical search for 
resonances can be performed in a tractable way using the period ratio 
$P^{\prime}/P$ as an initial guess.  

Eq. \ref{S4eq2} can be characterized in terms of mean motion and secular components.  When considering mean motion resonances, the $j_1-j_4$ 
coefficients are included such that they obey the zero sum condition, 
and the remaining terms are assumed to be zero.  The search for secular 
resonances usually consider only the condition where $j_3=-j_4=1$ and 
all other coefficients are zero.  The differences between these types 
of resonances manifests in the changes of the Keplerian orbital elements,
which define an orbit.  Further details on these changes can be found in 
\cite{Murray1999}, and their applicability of study within the Solar 
System.

Another method of searching for periodic orbits includes the 
characterization of chaos in the three-body problem.  Chaotic 
regions can alter the periodicity of the orbit, as well as the 
orientation of the orbit relative to some reference direction (i.e., 
precession of Eulerian angles).  \cite{Gozdziewski2013} and 
\cite{Migaszewski2012} demonstrated this principle with a 
chaos indicator (for details, see Section 4.5).

\subsection{Maximum Lyapunov exponent}

Time evolution of dynamical systems can be characterized through the use of 
the method of Lyapunov exponents \citep{Lyapunov1907}. First applications 
of these exponents to the general three-body problem were made by \cite
{Benettin1976,Benettin1980}; specific applications to the restricted 
three-body problem were first made by \cite{Jefferys1983}.  The method 
of Lyapunov exponents utilizes a description of two nearby trajectories 
in phase space for a third mass within the phase space.  The mathematical description follows a power law (usually linear) for stable orbits, 
whereas an exponential law is used for unstable orbits, with the 
Lyapunov exponent being a multiplicative power of such laws (i.e., 
$x^{\pm\lambda\:t},\:e^{\pm\lambda\:t}$).  

Most generally, the rate of divergence resulting from these laws is divided 
into 3 regimes.  A positive Lyapnuov exponent $\lambda$ denotes that a 
trajectory, corresponding to one degree of freedom, is diverging from the 
neighboring orbit, with the rate of divergence being characterized by a time 
series $\lambda(t)$.  A negative Lyapunov exponent implies that a trajectory 
is converging onto a stable manifold and can also be further described by a 
time series.  The third case considers a Lyapunov exponent that is exactly 
equal to zero, which means that the two trajectories are parallel to each 
other and by induction will remain so unless another force arises to 
disrupt the system.  The mathematical definition of the Lyapunov exponent 
\citep{Hilborn1994,Musielak2009} is 
\begin{equation}
\lambda = \lim_{t\rightarrow\infty} {1\over t - t_0} \ln\left({\delta(t) \over \delta(t_0)}\right)\ .
\label{S4eq3}
\end{equation}

Using the above basic description, one can calculate the spectrum or set of 
Lyapunov exponents for a given system with $n$ degrees of freedom in a $2n$ dimensional phase space.  Within the CR3BP, the system can be rotated with 
respect to the motion of the two large masses, so $n=3$ degrees of freedom 
exist within a 6 dimensional phase space.  Since in this example, the system 
is Hamiltonian, the trace of the Jacobian $J$ must be zero, which presents a 
symmetry among the Lyapunov exponents, with 3 positive and 3 negative exponents 
existing.  There are other examples for which such symmetry does not exist, 
and only 2 regimes, chaotic or dissipative, can be identified.  The sum of 
the exponents (${\rm Tr}\:J$) can then be positive or negative, and this 
sum determines whether a system is chaotic or dissipative, respectively.
  
The Jacobian for the $i$th mass is given by
\begin{align} \label{Sid_Jac}
J_i &= \begin{pmatrix}
	0 & 0 & 0& 1 & 0 & 0\\
	0 & 0 & 0& 0 & 1 & 0\\
	0 & 0 & 0& 0 & 0 & 1\\
	{\partial \ddot{\xi}_i \over \partial \xi_i} & {\partial 
	\ddot{\xi}_i \over \partial \eta_i} & {\partial \ddot{\xi}_i 
	\over \partial \zeta_i}& 0 & 0 & 0\\
	{\partial \ddot{\eta}_i \over \partial \xi_i} & {\partial 
	\ddot{\eta}_i \over \partial \eta_i} & {\partial 
	\ddot{\eta}_i \over \partial \zeta_i}& 0 & 0 & 0\\
	{\partial \ddot{\zeta}_i \over \partial \xi_i} & {\partial 
	\ddot{\zeta}_i \over \partial \eta_i} & {\partial 
	\ddot{\zeta}_i \over \partial \zeta_i}& 0 & 0 & 0\\
\end{pmatrix},
\end{align}
where the coordinates $\left(\xi,\eta,\zeta\right)$ denote an inertial 
reference frame.  The accelerations are given by Newton's Universal 
Law of Gravitation, i.e., $\textbf{a}_i = - \sum \limits_{i\neq j} 
{GM \over r_{ij}^3}
\hat{\textbf{r}}_{ij}$ and $r_{ij}^2 = (\xi_i - \xi_j)^2 + (\eta_i - 
\eta_j)^2 + (\zeta_i - \zeta_j)^2$.

\subsection{Fast Lyapunov Indicator}

Since the calculation of the Jacobian involves extra computational time, 
other methods have been developed to determine more efficiently the onset 
of chaos.  We discuss two such methods: the Fast Lyapunov Indicator (FLI) 
and the Mean Exponential Growth Factor of Nearby Orbits (MEGNO); the latter 
is discussed in the next section.  Using these methods chaos can be detected 
within 100,000 year timescales; note that previous methods required much 
longer timescales. 

The FLI is a method, which identifies two shortcomings of the previous 
approach of using the Lyapunov exponents, namely, normalization and 
dimensionality \citep{Froeschle1997}.  The problem of dimensionality 
is related to the computational time used to follow non-chaotic vectors 
within the phase space.  Thus, the supremum of the tangent vectors is 
used to determine the chaos indicator, with other vectors being neglected.  
This is shown by the following standard formula:
\begin{align}
{\rm FLI}(t) = {\rm sup}_j \left\|v_j(t)\right\|\ ,
\end{align}
where $j = 1\ldots m$ of a $m$-dimensional basis.  

The slope of the time series of FLI are used as indicators of chaos.  If the 
slope is steep and positive, then chaos is inferred; this is analogous to 
a positive Lyapunov exponent.  The slope may also remain flat, neither 
increasing or decreasing, which is the stable case.  Lastly, there is the possibility for a negative slope, but this is not generally realized in 
celestial mechanics as the systems under study are Hamiltonian and 
dissipation would imply a loss of energy.

\subsection{Mean exponential growth factor of nearby orbits}

An alternate technique for identification of chaos is the use of the 
MEGNO; this method is based on the maximum Lyapunov exponent but in 
a slightly different manner.  The method was developed by \citep
{Cincotta1999,Cincotta2000} as a general tool in probing chaos of 
Hamiltonian systems.  Since that time, it has been used to probe 
the dynamics of many different astrophysical systems \citep
{Gozdziewski2001,Gozdziewski2002a,Gozdziewski2002b,Laughlin2002,
PilatLohinger2003}.  For examples using the MEGNO method and 
other chaos indicator methods, we recommend \cite{Pavlov2003}, 
\cite{Kotoulas2004}, \cite{Barnes2006}, \cite{Funk2009}, 
\cite{Dvorak2010}, \cite{Hinse2010}, \cite{Mestre2011}, and 
\cite{Satyal2013}.

The MEGNO method employs the integral form of determining the maximum 
Lyapunov exponent.  In this sense, it can be interpreted as a measurement 
of the time-average mean value of the finite time Lyapunov exponent with 
the following definition \citep{Gozdziewski2001,Hinse2010}
\begin{equation}
Y(t) = {2 \over t} \int_0^t {\dot{\delta}(s) \over \delta(s)}s\:ds,
\label{S4eq4}
\end{equation}
along with its time-averaged mean value
\begin{equation}
\left\langle Y\right\rangle (t) = {1 \over t} \int_0^t Y(s)ds.
\label{S4eq5}
\end{equation}
The advantage of this representation is that $\left\langle Y\right\rangle (t)$ converges faster to its limit value even with the cost of two additional 
equations of motion \citep{Gozdziewski2001}.  This is due to the fact that 
the time-weighting factor amplifies the presence of stochastic behavior, 
which allows the early detection (i.e., shorter integration timescales) 
of chaos.  The additional differential equations are simply the differential 
forms of Eqs. (\ref{S4eq4}) and (\ref{S4eq5}) given respectively as
\begin{equation}
\dot{x} = {\dot{\delta} \over \delta}t \;\; {\rm and} \;\; \dot{w} 
= 2{x \over t},
\label{S4eq6}
\end{equation}
where $Y(t) = 2x(t)/t$ and $\left\langle Y\right\rangle (t) = w(t)/t$,
with all the variables having their usual meaning.

\section{Applications of the general three-body problem}

The general three-body problem has been considered in many 
astronomical and spaceflight settings, including galactic dynamics, 
stellar formation, and also the determination of trajectories for 
spacecraft missions, such as manned, satellite, and robotic landers.

\subsection{Astronomical settings}

Exploring the general three body problem within an astronomical 
setting can encompass a broad scale.  Here we discuss only the most basic 
cases.  In order for a three-body problem to be considered general, 
all three masses must interact with each other with no limitations on the eccentricity of their orbits.  One example is the case of triple stellar 
systems that are not hierarchical.  The basic dynamics of these systems 
can only be investigated numerically because a variable timestep method 
of integration (see Section 4) is required due to Kepler's 
Second Law.  The masses in such settings are likely to undergo several 
close approaches and great care must be taken during the point of close 
encounter between any of the masses. 

As previously discussed (see Section 3.4), the Pythagorean problem 
exemplifies the difficulty in obtaining meaningful results in the face 
of large sources of numerical error.  In terms of stellar dynamics, 
\cite{Reipurth2012} investigated this setup.  Examples include a finite 
sized cloud core around the three masses, which presents an even more 
intricate dynamical problem as the masses of the three bodies increase 
with time depending on the interaction with the cloud core.  These types 
of investigations provide an insight into the formation of wide binary 
stars ($>$ 100 AU separations) due to dynamical instabilities in the 
triple systems.

Conversely the general three body problem has been used to explain 
the formation of close binaries ($<$ 100 AU separations).  Results 
of numerical evaluation of orbits along with secular perturbations 
have explained the occurrence of close binaries \citep{Fabrycky2007}, 
who provided a specific mechanism for the formation of such systems 
through the Lidov-Kozai cycles \citep{Lidov1962,Kozai1962}; within 
these cycles large variations of eccentricity and inclination occur.  
The idea that stellar binary system may host planets has been explored 
in great detail \citep{Gonczi1981,Rabl1988,Dvorak1989,Holman1999,
PilatLohinger2002,PilatLohinger2003,Dvorak2004,Musielak2005,
PilatLohinger2007,Haghighipour2010,Cuntz2014}, with the additional 
observational verification by the NASA Kepler mission (see Section 1).    
The consideration of binaries hosting planets will be addressed in 
more detail in Sections 8 within different regimes of approximation.

\subsection{Exoplanets}

Recently, using the {\it Kepler} data, over 700 exoplanet candidates 
have obtained the status of exoplanets based on a statistical method 
to identify the likelihood of false positive candidates among the 
sample \citep{Lissauer2014,Rowe2014}.  This portends to the general 
three body problem as each considered case includes the possible 
interaction of three masses.  This is due to the possible geometric 
configurations in the photometric method.  Some examples include the 
presence of third light contamination, background eclipsing binaries, 
and stellar triple systems.  Prior to this major identification of 
exoplanets, a method called transit timing variations (TTVs) has been 
used to identify systems with multiple planets (Kepler-11, Kepler-36, 
Kepler-9); the method determines how three-body interactions would 
affect the timing of transits relative to a linear prediction 
\citep{Lissauer2011,Carter2012,Holman2010}.  We present the basics 
of TTVs and defer the reader to a more in depth description by 
\cite{Winn2011}.

The method of TTVs relies on the observable deviation from Keplerian 
dynamics.  The deviation reveals the existence of additional bodies, 
which in principle is a similar method that was used in the discovery 
of Neptune through variations in the orbit of Uranus \citep{Adams1846,
Airy1846,Challis1846,Galle1846}.  Using the times of transit and the 
planetary period of a transiting planet, a fitting is made to a linear 
function to predict the subsequent eclipses ($t_n = t_o + nP$) of that 
planet.  When the planet eclipses a host star early or late relative to 
the predicted time of the fitted function, a TTV is observed.  The power 
of this method was illustrated by \cite{Nesvorny2012} with the detection 
of a non-transiting planet (KOI-872c) inferred by observations of the 
seen transiting planet (KOI-872b).  The vetting of this planet underwent 
intense scrutiny from the perspective of the three-body problem as only 
particular solutions allowed for the determination of the planet's mass 
as well as keeping the perturbing planet stable.  

Many other discoveries have been made detecting multiple planet systems, 
including KOI-142 which is another discovery of a planet pair based 
on their mutual interactions \citep{Nesvorny2013}.  Much previous work 
was related to predicting the degree to which discoveries could be made 
based on the possible perturbations, and involved determining the 
sensitivity required to perform observations \citep{Miralda2002,
Holman2005, Agol2005}.  TTVs 
along with transit duration variations (TDVs) have even been suggested 
as possible methods to detect the presence of extrasolar moons \citep
{Kipping2009}, although none so far have yet to be observed.  Additional 
discussion on the requirements for the detection of extrasolar moons is 
presented in $\S$\ref{sec:exomoon}.

To account for exoplanets and exomoons in extra-solar planetary systems 
with the number of objects higher than three, additional new tools, such 
as {\it TTVfast} and {\it photodynam}, have been developed with full 
n-body capabilities built-in \citep{Deck2014,Carter2011,Pal2012}.

\subsection{Spacecraft trajectories}

Spaceflight applications involve the determination of spacecraft 
trajectories for satellite and lander space missions.  NASA's 
{\it Apollo 8 - 10} missions to the Moon required accurate numerical 
solutions to the three-body problem.  NASA also launched several 
discovery missions, with the most notable being the {\it Hubble 
Space Telescope, CHANDRA X-Ray Observatory, SPITZER Space Telescope, 
Kepler Space Telescope}, and the upcoming {\it James Webb Space 
Telescope}.  Each of these missions involve different solutions to 
three-body problem and all but {\it Kepler} orbit the Earth.  One 
of the most interesting satellites will be the {\it Transiting 
Exoplanet Survey Satellite} (TESS{\footnote{http://tess.mit.edu/}}) 
as it will be in a high-Earth orbit and in a 2:1 resonance with 
the Moon.  For these applications, it is required to consider the 
motions of the terrestrial objects relative to each other, including 
the motions of the Moon and Mars.  The orbits of both of these objects 
can be affected by the perturbations from other planets. 

More basic applications of the general three body problem involve 
artificial satellites commonly used for communication and/or military 
purposes.  These satellites are generally considered within the realm 
of a restricted three-body problem (see Sections 6, 7 and 8).   A 
restricted three body problem simplifies the solution because one of 
the masses is much less than the others and has negligible gravitational 
influence.  Thus, the much larger masses have approximately Keplerian 
orbits.  This is the case for the Earth-Moon-satellite system, and it 
is discussed in detail in Section 8.

\section{The circular restricted three-body problem}

\subsection{Governing equations}

Based on the criteria stated in Section 1, the circular restricted
three-body problem (CR3BP) requires that the two objects have their masses 
significantly larger than the third one ($M_1 >> M_3$ and $M_2 >> M_3$), 
and that the motions of $M_1$ and $M_2$ are limited to circular orbits around 
their center of mass.  In the literature devoted to the CR3BP, $M_1$ and 
$M_2$ are typically referred to as the primaries \citep{Szebehely1967},
but in stellar dynamics $M_1$ and $M_2$ are called the primary and secondary, 
respectively, with the assumption that $M_1 \geq M_2$.  In this paper, we 
use the former nomenclature when a clear distinction between $M_1$ and $M_2$ 
is not necessary; however, we use the second nomenclature when a distinction 
between $M_1$ and $M_2$ becomes important in describing a specific astronomical 
application.   
 
Now, if the third mass moves in 3D, this case is called the 3D CR3BP; 
if it moves in the same plane as the primaries, we call this case the  
planar CR3BP.  Since the gravitational influence of the third body on the 
primaries is negligible, the orbits of the primaries are described by the 
two-body problem, whose solutions are well-known.  Once the solutions for 
the primaries are known, they can be used to determine the motion of the 
third body resulting from the gravitational field of the primaries.  From 
a mathematical point of view, the problem becomes simpler than the {\it 
general} three-body problem.  Nevertheless, for some sets of initial 
conditions the resulting motions of the third body remain unpredictable. 

\begin{figure}
\centering
\begin{tabular}{cc}
\includegraphics[width=0.4\linewidth]{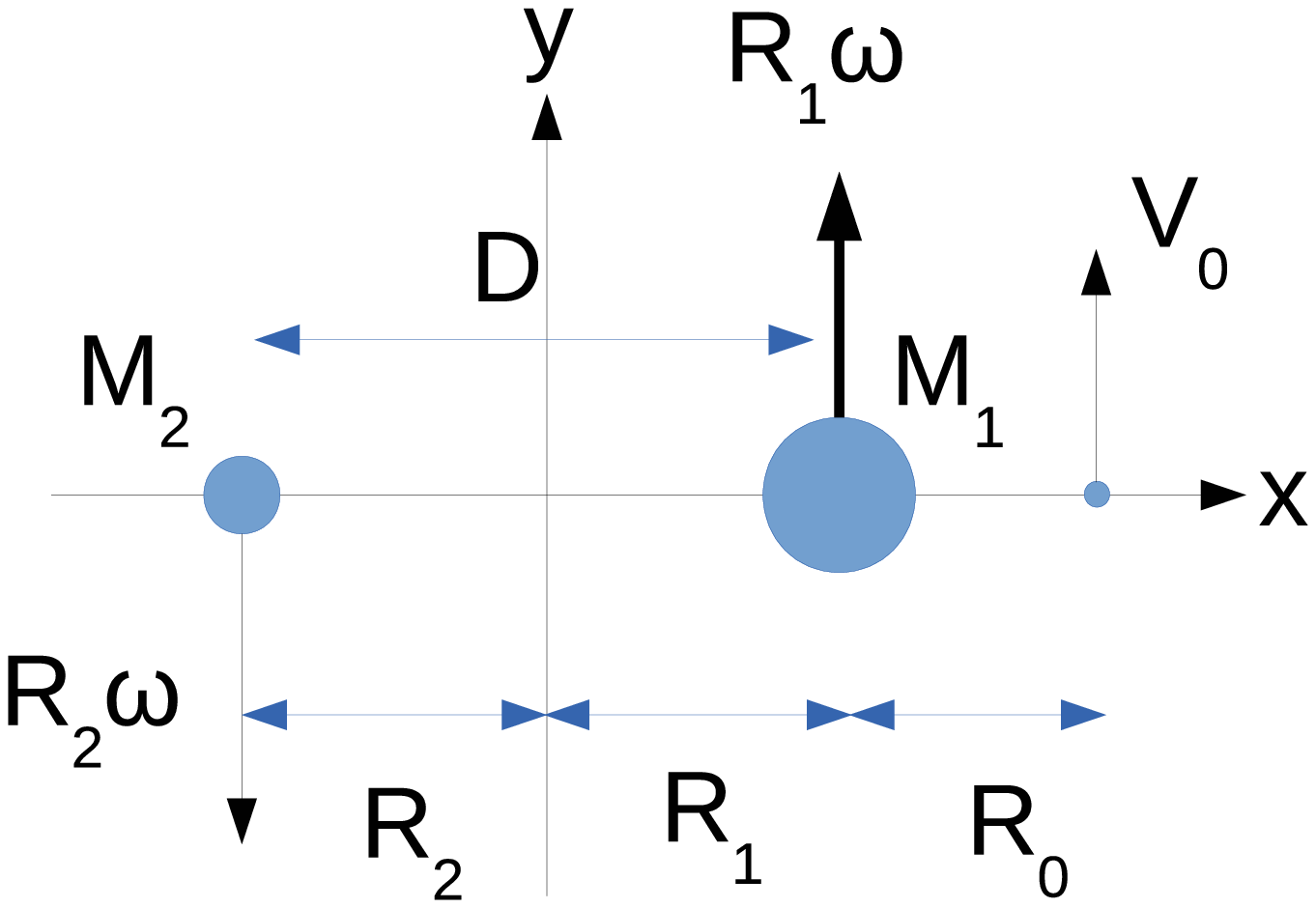}&
\includegraphics[width=0.4\linewidth]{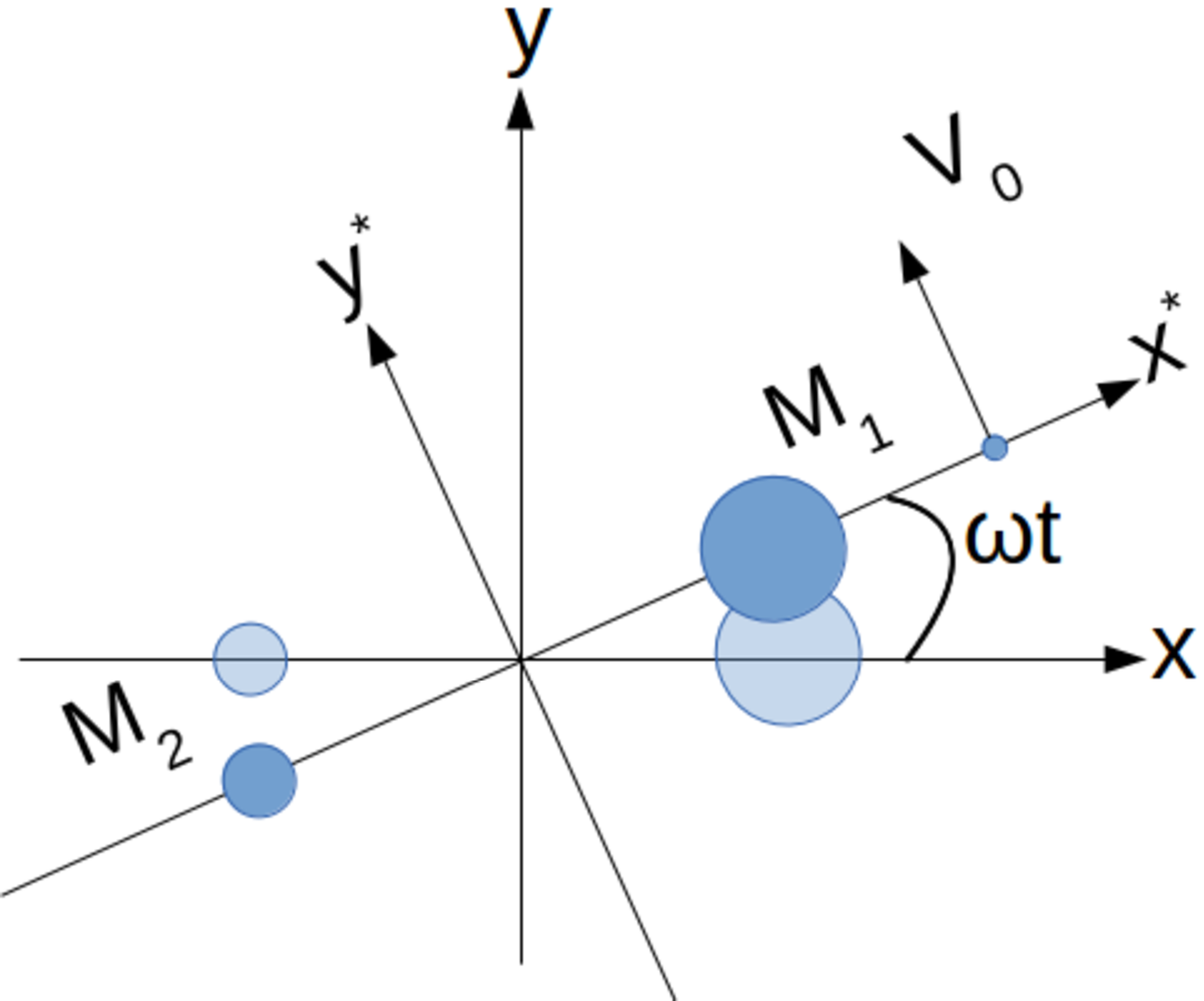}
\end{tabular}
\caption{Non-rotating and rotating coordinate systems used 
to describe the CR3BP.  Reproduced after \cite{Eberle2008}.
Copyright 2008 Astronomy \& Astrophysics.}   

\label{fig:Coordinates}
\end{figure}

Let us choose the origin of a non-rotating coordinate system (see Fig. 
\ref{fig:Coordinates}) to be at the center of mass of the primaries, so 
we can write $M_1 \vec R_1 = M_2 \vec R_2$, where $M_1$ and $M_2$ are the 
masses of the primaries with $\vec R_1$ and $\vec R_2$ denoting their position 
vectors with respect to the origin of the coordinate system.  To describe 
the problem mathematically, we follow \cite{Szebehely1967}, \cite{Danby1988} 
and \cite{Roy2005}. 

The components of the position vectors are: $\vec R_1 = (X_1, Y_1, Z_1)$,
$\vec R_2 = (X_2, Y_2, Z_2)$ and $\vec R_3 = (X_3, Y_3, Z_3)$, and they 
are used to express $\vec r_{31} = \vec R_1 - \vec R_3 \equiv \vec r_1$
and $\vec r_{32} = \vec R_2 - \vec R_3 \equiv \vec r_2$ in terms of their
components.  The equations of motion for the CR3BP can be written as
\begin{equation}
{{d^2 \vec R_i} \over {dt^2}} = G \sum_{j=1}^{2} {{M_j} 
\over {r_{j}^3}} (\vec R_j - \vec R_3)\ ,
\label{S6eq1}
\end{equation}

\noindent
where $r_j = [(X_j - X_3)^2 + (Y_j - Y_3)^2 + (Z_j - Z_3)^2]^{1/2}$.

We define $M = M_1 + M_2$, $\mu = M_2 / M$ and $\alpha = 1 - \mu$, which 
give $M_1 = \alpha M$ and $M_2 = \mu M$.  With the gravitational force
being equal to the centripetal force, we have 
\begin{equation}
V_i^2 = R_j {{G M_{3-j}} \over {D^2}} = R_j^2 \omega^2\ ,
\label{S6eq2}
\end{equation}

\noindent
where $D = R_1 + R_2$, $\omega$ is the angular frequency or mean motion, 
and $j = 1$ and $2$.  Since $M_{3-j} = M R_j / D$, Kepler's third law 
$\omega^2 = G M / D^3$ is automatically obtained.  Moreover, we also 
have $R_1 = \mu D$ and $R_2 = \alpha D$. 
 
Now, once the large two masses are in circular orbits, we write:
$X_1 (t) = \mu D\ \cos \omega t$, $Y_1 (t) = \mu D\ \sin \omega t$,
$X_2 (t) = - \alpha D\ \cos \omega t$ and $Y_2 (t) = - \alpha D\ \sin 
\omega t$, with $Z_1 = Z_2 = 0$, if the orbits are in the same plane.

Introducing $X_3 = D x$, $Y_3 = D y$ and $Z_3 = D z$, and allowing the
third body to move in 3D, we use Eq. (\ref{S6eq1}) to derive the following
set of equations of motion
\begin{equation}
\ddot x = - {{\alpha} \over {r_1^3}} (x - \mu \cos \tau)
- {{\mu} \over {r_2^3}} (x + \alpha \cos \tau)\ ,
\label{S6eq3a}
\end{equation}
\begin{equation}
\ddot y = - {{\alpha} \over {r_1^3}} (y - \mu \sin \tau)
- {{\mu} \over {r_2^3}} (y + \alpha \sin \tau)\ ,
\label{S6eq3b}
\end{equation}
and
\begin{equation}
\ddot z = - \left ({{\alpha} \over {r_1^3}} + 
{{\mu} \over {r_2^3}} \right )\ z\ ,
\label{S6eq3c}
\end{equation}

\noindent
where $\tau = \omega t$, and $\ddot x$, $\ddot y$ and $\ddot z$ 
represent the second-derivative of $x$, $y$ and $z$ with respect 
to $\tau$, respectively.  This set of equations describes the 
CR3BP in the non-rotating coordinate system.

We write the above set of equations in the rotating (synodic) 
coordinate system by using the following relationships between 
the coordinates $x$, $y$ and $z$ in the inertial system, and 
the coordinates $x^*$, $y^*$ and $z^*$ in the synodic system: 
$x = x^* \cos \tau - y^* \sin \tau$, $y = x^* \sin \tau + y^* 
\cos \tau$ and $z = z^*$.  Therefore, the set of equations 
describing the CR3BP in the synodic coordinate system can 
be written as \citep{Eberle2010c} 
\begin{equation}
\ddot x^* - 2 \dot y^* = x^* - {{\alpha} \over {r_1^3}} (x^* - \mu) - 
{{\mu} \over {r_2^3}} (x^* + \alpha)\ ,
\label{S6eq4a}
\end{equation}
\begin{equation}
\ddot y^* + 2 \dot x^* = \left ( 1 - {{\alpha} \over {r_1^3}} - 
{{\mu} \over {r_2^3}} \right )\ y^*\ ,
\label{S6eq4b}
\end{equation}
and
\begin{equation}
\ddot z^* = - \left ({{\alpha} \over {r_1^3}} + {{\mu} \over 
{r_2^3}} \right )\ z^*\ ,
\label{S6eq4c}
\end{equation}

\noindent
where $r_1 = D [(x^* - \mu)^2 + (y^*)^2 + (z^*)^2]^{1/2}$, 
and $r_2 = D [(x^* + \alpha)^2 + (y^*)^2 + (z^*)^2]^{1/2}$. 

\subsection{Lagrange points}

\cite{Lagrange1772} found interesting solutions to the CR3BP that 
describe equilibrium positions of the third body when all net forces 
acting on it are zero.  In this case, Eqs (\ref{S6eq4a}), (\ref{S6eq4b}),
and (\ref{S6eq4c}) reduce to 
\begin{equation}
x^* - {{\alpha} \over {r_1^3}} (x^* - \mu) - {{\mu} \over {r_2^3}} 
(x^* + \alpha) = 0\ ,
\label{S6eq5a}
\end{equation}
\begin{equation}
y^* \left ( 1 - {{\alpha} \over {r_1^3}} - {{\mu} \over {r_2^3}} 
\right ) = 0\ ,
\label{S6eq5b}
\end{equation}
and
\begin{equation}
z^* \left ({{\alpha} \over {r_1^3}} + {{\mu} \over {r_2^3}} \right )
= 0\ .
\label{S6eq5c}
\end{equation}

Taking $z^* \neq 0$ gives $y^* = 0$; however, $z^* = 0$ gives $y^* \neq 0$,
which means that the equilibrium solutions are confined to a plane.  We take 
this plane to be the $x^*y^*$-plane.  Thus, with $y^* \neq 0$, we obtain 
$r_1 = r_2 = 1$, which shows that the third body can be at either of these
two points located at the same distance from the primaries as the two primaries 
are from each other \citep{Roy2005}.  Clearly, an equilateral triangle is formed 
and the two equilibrium points are called the Lagrange triangular $L_4$ and $L_5$ 
points (see Fig. 6).  Both points are stable for the mass ratios $0 \leq \mu \leq 
\mu_0$, where $\mu_0 = (1 - \sqrt{69} / 9) / 2$ is called Routh's critical 
mass ratio \citep{Bardin2002}.  Now, with $y^* = z^* = 0$, it is easy to show 
that Eq. (\ref{S6eq5a}) has three equilibrium solutions located on the line 
passing through the primaries.  These three solutions are called the Lagrange 
collinear $L_1$, $L_2$ and $L_3$ points (see Fig. 6), and they are unstable 
for any value of $\mu$ \citep{Ragos2000}.  The Lagrange points are also called 
the libration points.  

\subsection{Jacobi's integral and constant}

Let us introduce a potential function $\phi^*$ defined as
\begin{equation}
\phi^* = {1 \over 2} \left [ (x^*)^2 + (y^*)^2 \right ] + 
{{\alpha} \over {r_1}} + {{\mu} \over {r_2}}\ ,
\label{S6eq6}
\end{equation}

\noindent
and write Eqs. (\ref{S6eq4a}), (\ref{S6eq4b}) and (\ref{S6eq4c}) 
in terms of $\phi^*$,

\begin{equation}
\ddot x^* - 2 \dot y^* = {{\partial \phi^*} \over {\partial x^*}}\ , 
\label{S6eq7a}
\end{equation}
\begin{equation}
\ddot y^* + 2 \dot x^* = {{\partial \phi^*} \over {\partial y^*}}\ ,
\label{S6eq7b}
\end{equation}
and
\begin{equation}
\ddot z^* = {{\partial \phi^*} \over {\partial z^*}}\ .
\label{S6eq7c}
\end{equation}

We multiply Eqs. (\ref{S6eq7a}), (\ref{S6eq7b}) and (\ref{S6eq7c}) by 
$\dot x^*$, $\dot y^*$ and $\dot z^*$, respectively, sum them up, and
integrate them to obtain the following Jacobi integral 
\begin{equation}
(v^*)^2 = 2 \phi^* - C^*\ ,
\label{S6eq8}
\end{equation}

\noindent
where $(v^*)^2 = (\dot x^*)^2 + (\dot y^*)^2 + (\dot z^*)^2$, and $C^*$
is the Jacobi constant in the rotating coordinate system given by
\begin{equation}
C^* = \mu + 2 \mu \rho_0 + {{1 - \mu} \over {\rho_0}} + {{1\mu} \over 
{1 + \rho_0}} + 2 \sqrt{\rho_0 (1 - \mu)}\ .
\label{S6eq8a}
\end{equation}

\noindent
Here $\rho_0 = R_0 / D$ (see Fig. 5) represents the normalized initial 
position of the third mass.   

Having obtained the Jacobi integral in the rotating coordinate system,
we also write its explicit form in the non-rotating coordinate system 
\begin{equation}
v^2 - 2 (x \dot y - y \dot x) = 2 \left ( {{\alpha} \over {r_1}} + 
{{\mu} \over {r_2}} \right ) - C\ ,
\label{S6eq9}
\end{equation}

\noindent
where $v^2 = \dot x^2 + \dot y^2 + \dot z^2$, and $C$ is the Jacobi 
constant in the non-rotating coordinate system.

\subsection{Zero velocity curves} 

In the case when the particle's velocity $v^*$ is zero, the Jacobi integral 
given by Eq. (\ref{S6eq8}) reduces to $2 \phi^* = C^*$, which defines the
zero velocity curves (ZVCs) for the third body.  Once $C^*$ is determined 
by the initial conditions, the ZVC encloses a region in space where $2 
\phi^* > C^*$ and where the third body must be found.  The third body 
cannot be outside of this region because its velocity would be imaginary 
since $2 \phi^* < C^*$.  The ZVCs do not give any information about orbits 
of the third body in the region of space where it is allowed to move. 

\begin{figure}
\centering
\begin{tabular}{ccc}
	\includegraphics[width = 0.33\linewidth]{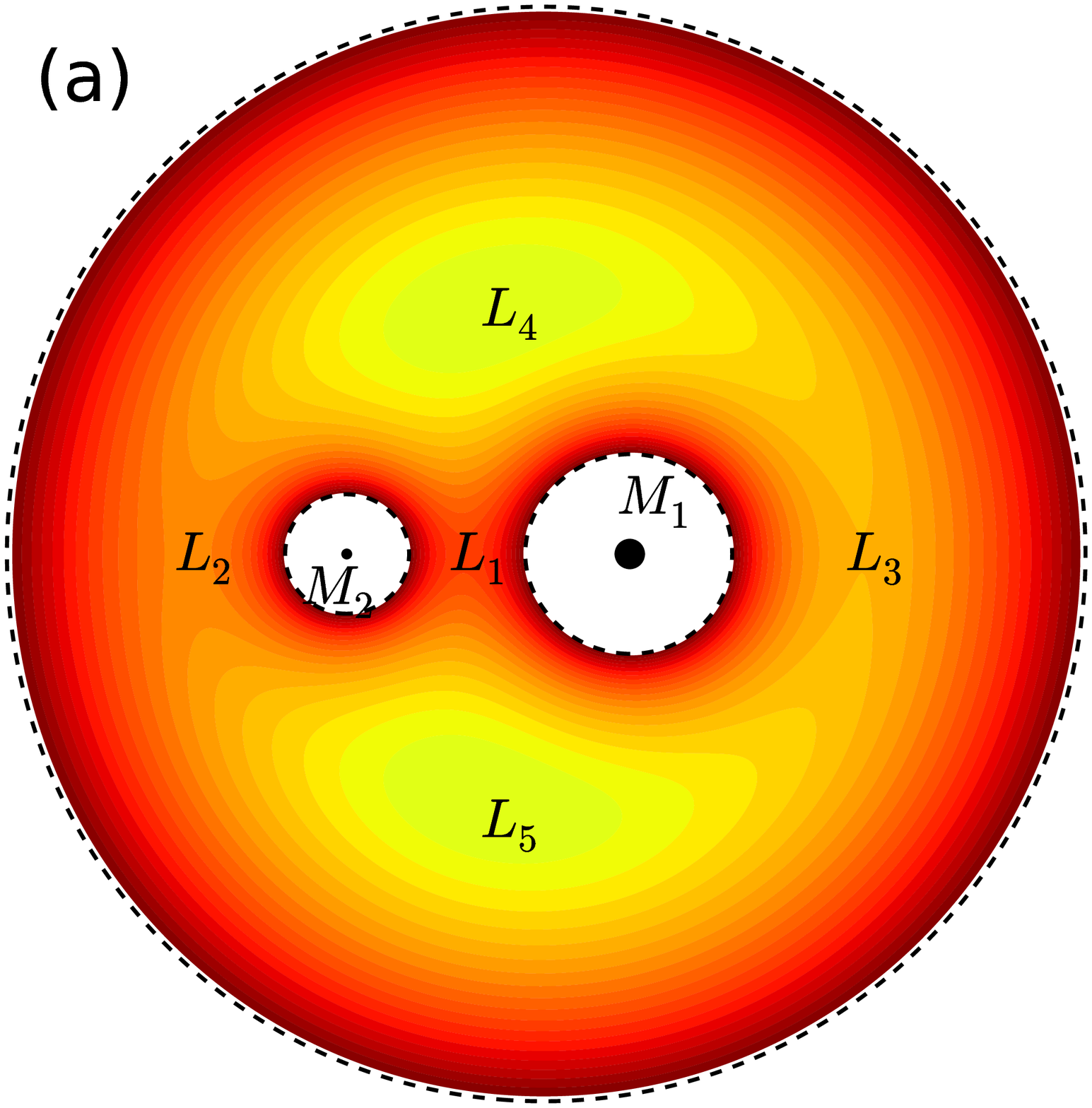}&
	\includegraphics[width = 0.33\linewidth]{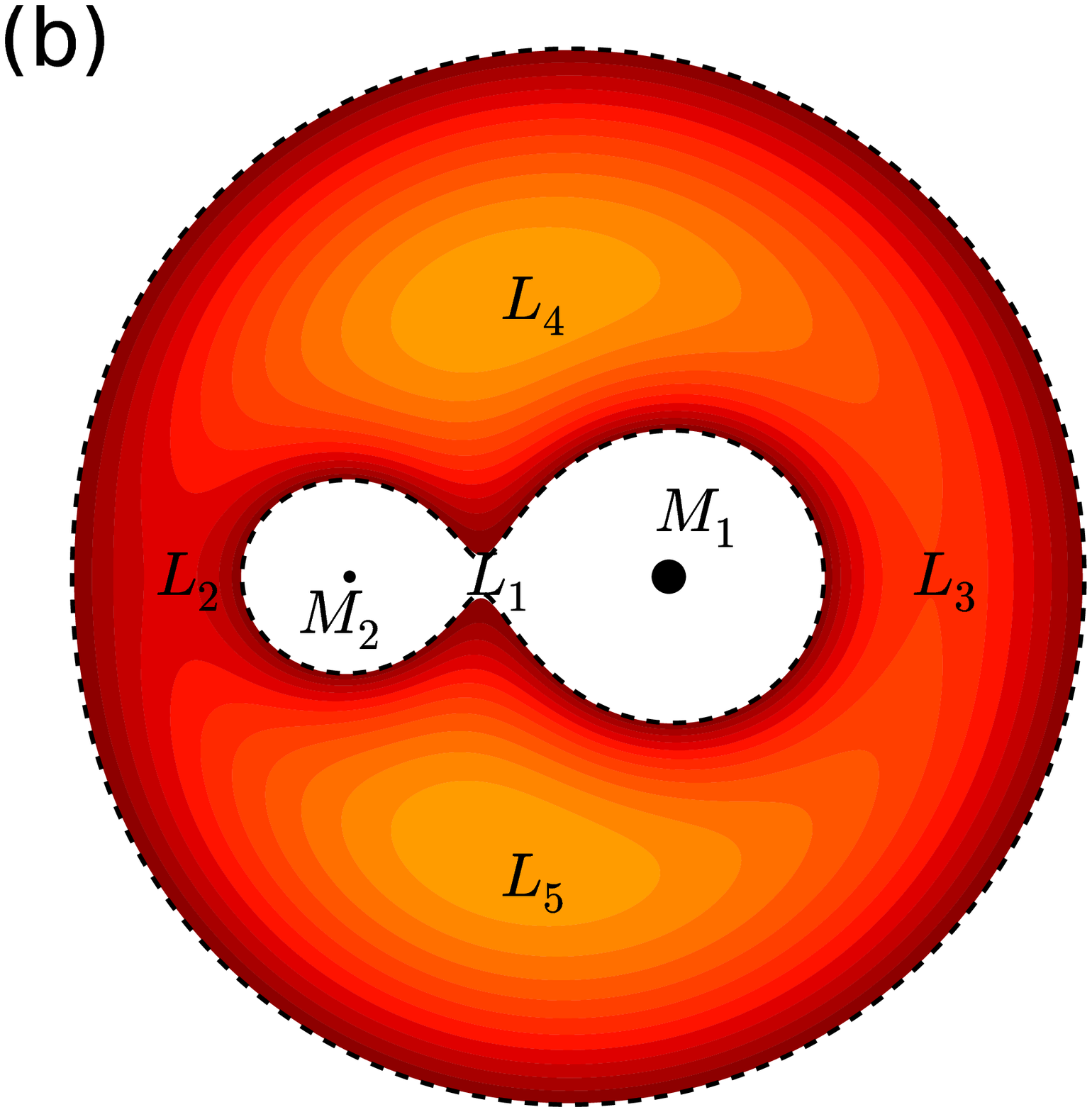}&
	\includegraphics[width = 0.33\linewidth]{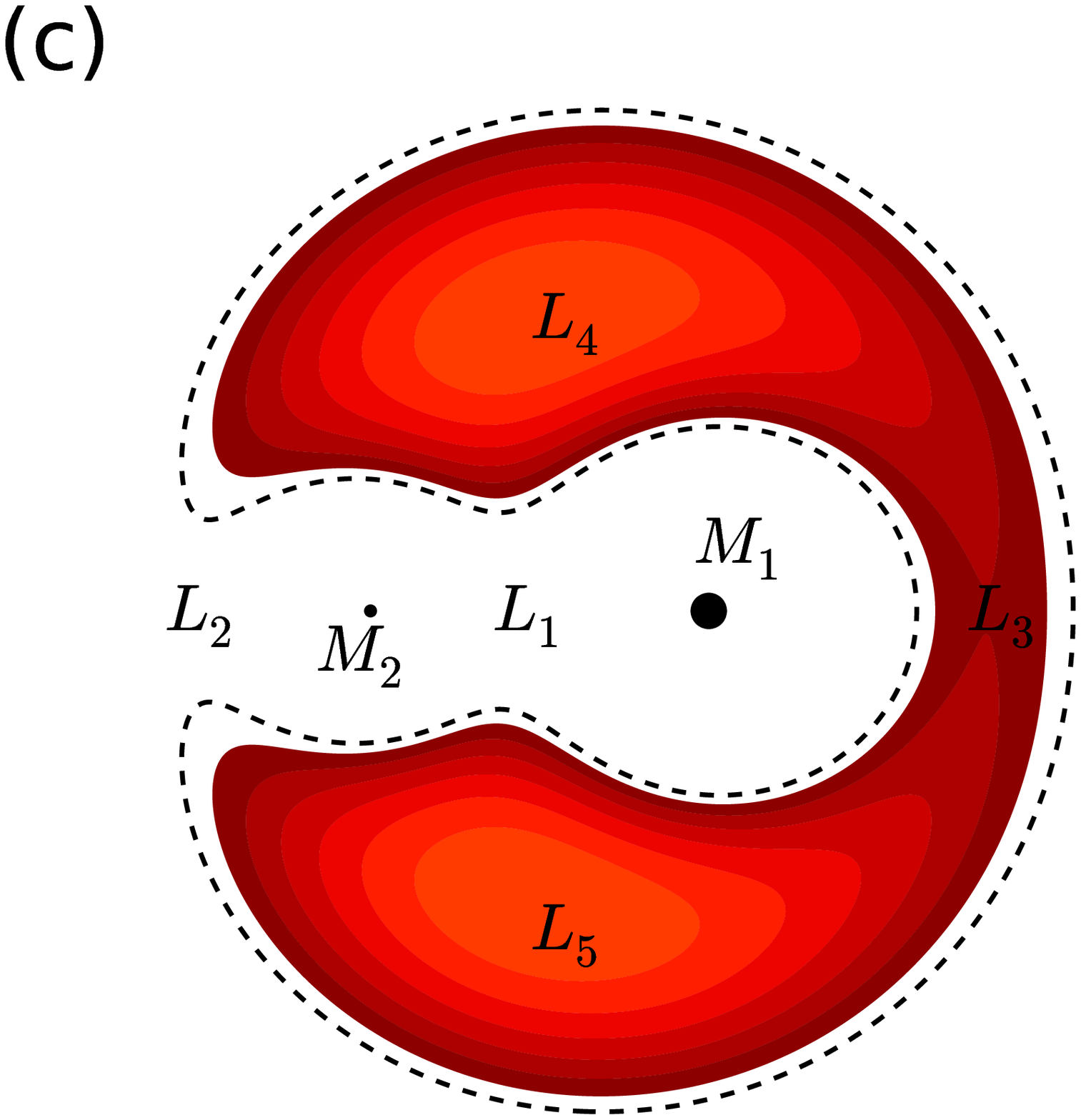}\\
	\includegraphics[width = 0.33\linewidth]{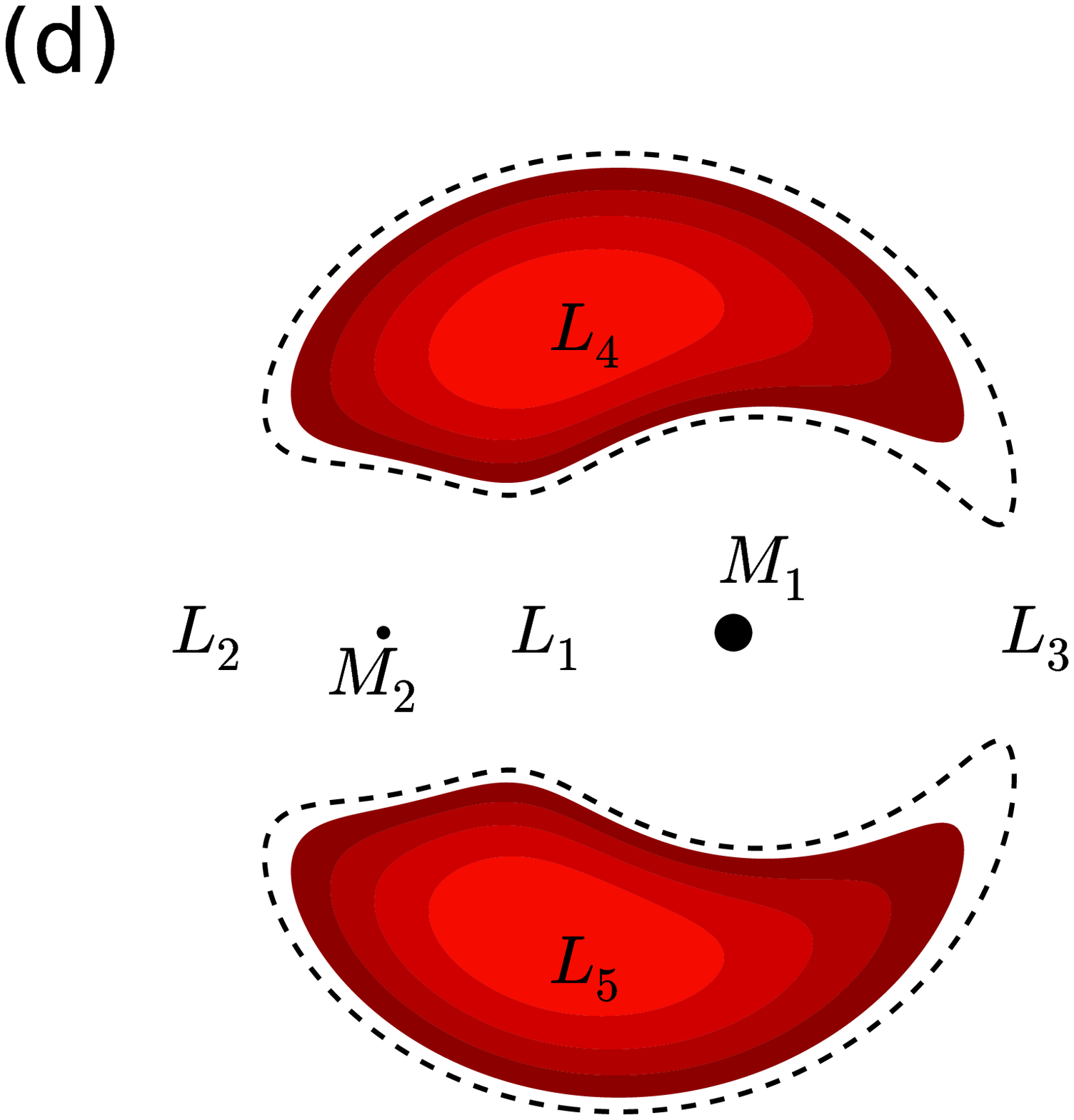}&
	\includegraphics[width = 0.33\linewidth]{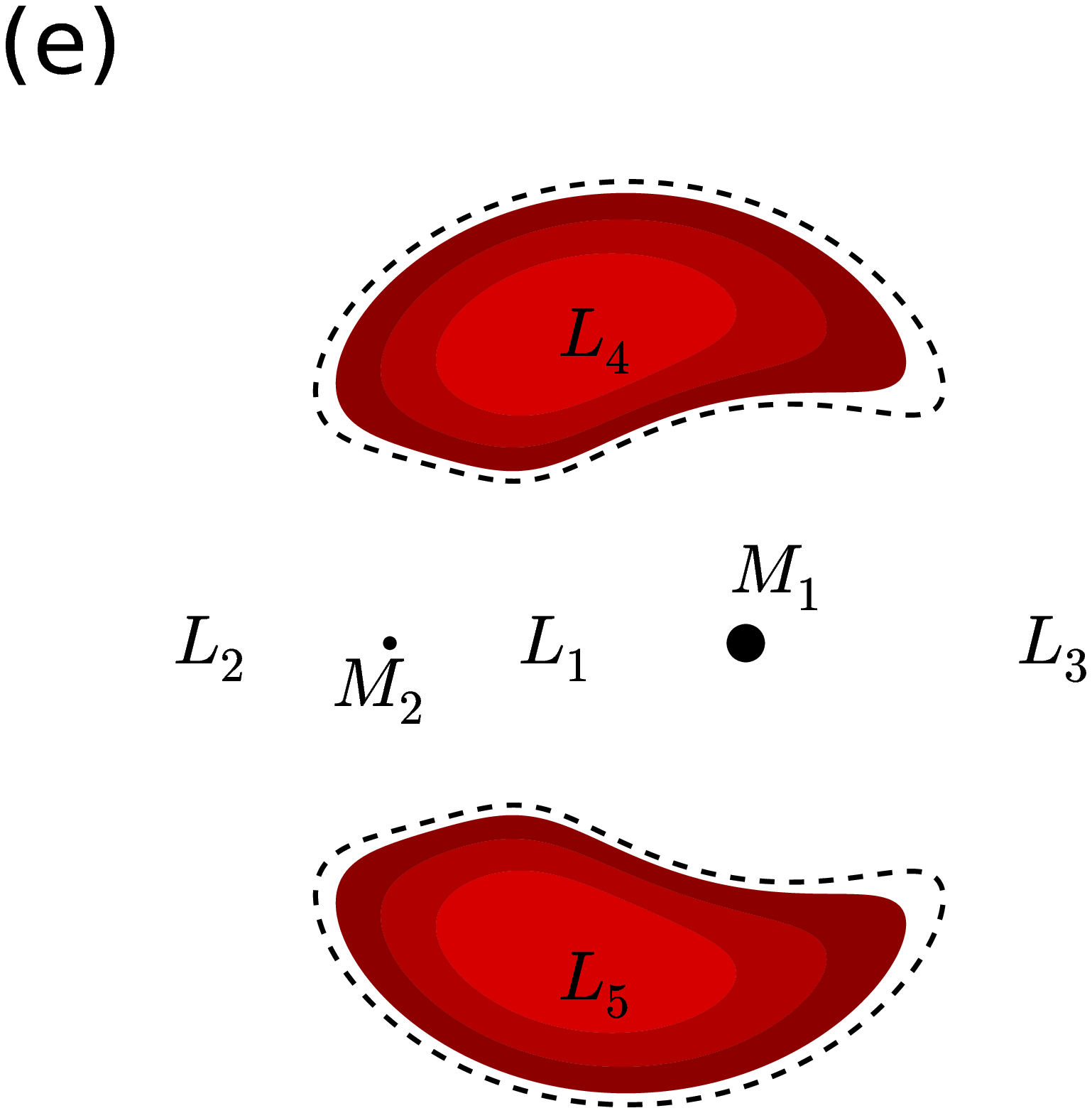}&
	\includegraphics[width = 0.33\linewidth]{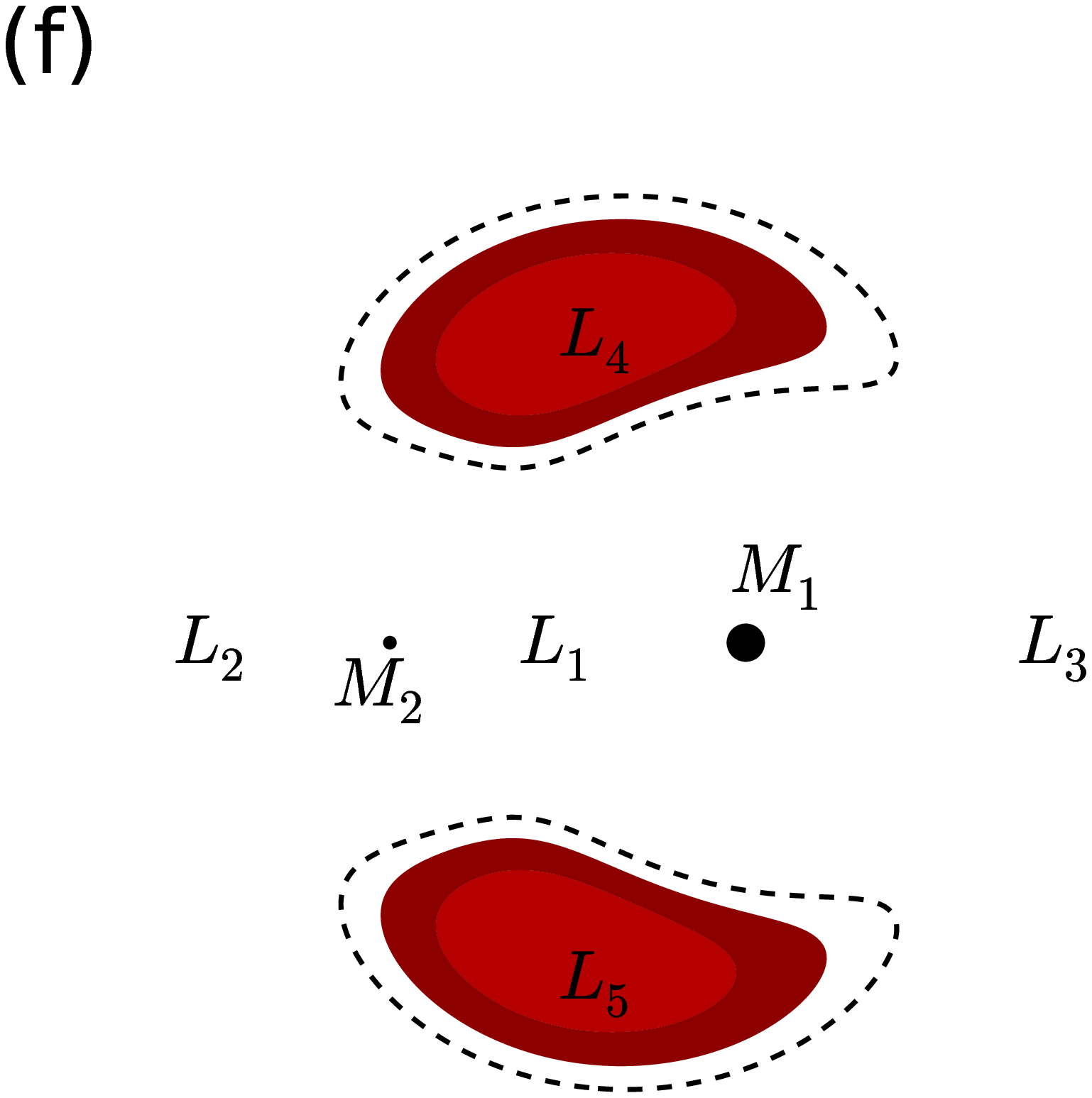}\\
\end{tabular}
\caption{Zero velocity curves in the synodic coordinate system,
computed for a mass ratio $\mu$ = 0.3 and various initial positions 
$\rho_o = R_o / D$ of the third mass.  The dashed lines represent 
the boundary of zero velocity while the coloured region indicates 
the forbidden zones. 
(a) $\rho_o$ = 0.2; (b) $\rho_o$ = 0.28; (c) $\rho_o$ = 0.36; 
(d) $\rho_o$ = 0.44; (e) $\rho_o$ = 0.52; (f) $\rho_o$ = 0.6; }
\label{fig:ZVC}
\end{figure}

The shape of the ZVCs is uniquely determined by the values of $C^*$.
Changes in the shape of the ZVCs in the $x^*y^*$ plane with changing 
$C^*$ expressed in terms of the initial position of the third mass 
$\rho_0$ (see Eq. \ref{S6eq8a}) are plotted in Fig. \ref{fig:ZVC}; 
the region inaccessible to the third mass is colored.  The presented 
results were obtained for $\mu = 0.3$ and $\rho_0$ ranging from $0.2$, 
which corresponds to a large value of $C^*$, to $0.6$, which corresponds 
to a small value of $C^*$.  Note that the regions inaccessible to the 
third mass shrink and eventually vanish at the Lagrange points $L_4$ 
and $L_5$, as shown in Figs. \ref{fig:ZVC}d,e,f.        

Transit orbits in the CR3BP are the trajectories that pass through the 
neck region of the ZVCs, which occurs in the vicinity of the Lagrange 
points $L_1$, $L_2$, and $L_3$ where the ZVCs open at these libration 
points at certain critical values of $C^*$ (see Fig. \ref{fig:ZVC}).  
Moving along these transit orbits, the third body can transit between 
the primaries or even escape from the system.  Extensive numerical 
studies of transit orbits in the CR3BP were recently performed by 
\cite{Ren2012}.  They established the necessary and sufficient 
conditions for such transition by using manifolds of the vertical 
and horizontal Lyapunov orbits along with the transit cones, and 
applied the results to the Sun-Earth CR3BP. 

\subsection{Periodic and quasi-periodic orbits}

\cite{Poincare1892} attached great importance to periodic orbits in the 
CR3BP, and he used his method (see Section 3.2) to identify and classify 
periodic motions of the third body with respect to the synodic coordinate 
system.  He searched for periodic orbits by taking $\mu = 0$ and then used 
analytical continuation to find periodic orbits for $\mu > 0$.  In his 
classification of periodic orbits, {\it premi\'ere sorte} contains the 
orbits generated from the known two-body circular orbits; {\it deuxi\'eme 
sorte} contains the orbits generated from the known two-body elliptical 
orbits; and {\it troisi\'eme sorte} contains the orbits generated from 
the known two-body circular orbits, but with a non-zero inclination of 
the third body with respect to the plane in which the orbits of the 
primaries are confined. 
  
Poincar\'e's work on periodic orbit was continued by \cite{Darwin1897,
Darwin1909}, \cite{Moulton1920} and \cite{Stromgren1922}.  Then, 
\cite{Stromgren1935} performed extensive studies of the periodic orbits 
for $\mu = 0.5$ and demonstrated that they belong to the {\it premi\'ere 
sorte}.  He concluded that the termination of this family of orbits must 
be an asymptotic periodic orbit spiraling into the Lagrange points $L_4$ 
and $L_5$; a proof of this conjecture was given by \cite{Henrard1973} 
and \cite{Buffoni1999}.  

Various families of periodic orbits in the commensurable 3D CR3BP, 
where two bodies are close to commensurability in their mean motions around 
the central body, were identified and classified by \cite{Sinclair1970}.  
Moreover, a special class of periodic orbits in the CR3BP resulting from 
an analytical continuation of Keplerian rectilinear periodic motions was 
discovered by \cite{Kurcheeva1973}, who showed that the period of such 
orbits is an analytical function of $\mu$; the conditions when these 
orbits are stable were established analytically by \cite{Ahmad1995}.
Comprehensive studies of the periodic orbits in the CR3BP were performed 
by \cite{Henon1965,Henon1974} who classified them and also investigated 
intersections between families of generating orbits \citep{Henon1997,
Henon2001}. 

There are periodic orbits around the Lagrange points as demonstrated by 
\cite{Stromgren1935}.  Periodic orbits in the vicinity of these points 
are typically classified as halo, vertical, and horizontal Lyapunov orbits 
\citep{Henrard2004}.  They arise either as a continuation of infinitesimal 
oscillations or as bifurcations of the planar periodic orbits around the points.  
Some unstable periodic orbits around $L_1$, $L_2$ and $L_3$ were identified and 
classified by \cite{Zagouras1979} for $\mu = 0.00095$ (Sun-Jupiter).  Since $L_4$ 
and $L_5$ are stable \citep{Szebehely1967,Roy2005}, there are stable zones around these points for certain physical conditions \citep{Dvorak1991}.  The existence 
of periodic orbits in the vicinity of the $L_4$ and $L_5$ have been studied.  
\cite{Perdios1991} identified the long and short periodic orbits near the two libration points for the mass parameter $\mu$ ranging from $0.03$ to $0.5$, 
and \cite{Markellos1991} found families of remarkable termination orbits.  
According to Routh's criterion, the existence of periodic orbits near the 
triangular libration points is limited to the mass ratios $0 \leq \mu < \mu_0$, 
where $\mu_0$ is Routh's critical mass ratio \citep{Bardin2002}.  Moreover, 
\cite{Broucke1999} adopted canonical units to define the unit circle and 
formulated a symmetry theorem that was used to discover new long and 
short-period orbits around $L_4$ and $L_5$.  

All orbits discovered in the above studies were symmetric periodic orbits, 
which are much more easy to generate than non-symmetric periodic orbits.  
Systematic explorations of non-symmetric periodic orbits in the 3D 
CR3BP were done by \cite{Zagouras1977} and \cite{Zagouras1996}, who 
discovered new orbits and classified them.  Searches for non-symmetric 
periodic orbits near $L_4$ (because of the symmetry conditions the orbits 
are the same at $L_5$) were performed by \cite{Henrard1970,Henrard2002}, 
who discovered an infinite number of families of non-symmetric periodic 
orbits in the vicinity of the points.  In recent work of \cite{Henrard2004} 
showed that some of these orbits are emanating from homoclinic orbits (see 
Section 6.7).

The existence of quasi-periodic orbits around the libration points was 
confirmed by using different semi-analytical, numerical, and combined 
analytical-numerical methods \citep{Ragos1997,Jorba1999,Gomez1999}.  
An overview of quasi-periodic orbits around $L_2$ is given by \cite{
Henrard2004}, who showed that different families of such orbits exist 
and that these families are associated with the known halo, vertical 
and horizontal Lyapunov orbits.  They also presented a fast numerical 
method based on multiple Poincar\'e sections (see Section 6.7) and 
used it to find quasi-halo periodic orbits.  \cite{Henrard2004} 
demonstrated that the method gives full convergence for a given 
family of quasi-periodic orbits, and that it is robust near chaotic 
regions.

\subsection{Circular Hill problem}

In the Hill three-body problem, the mass of one primary dominates over 
the other masses ($M_1 >> M_2$), and the mass of the third body is 
negligible when compared to both primaries ($M_1 >> M_3$ and $M_2 >> M_3$).  
Because of these mass relationships, $M_2$ moves around $M_1$, and if its 
orbit is circular, the problem is called the {\it circular} Hill problem.  
The third body can move either in 3D space or in the same plane as the 
circular orbit, and its orbit is determined by the equations of motion 
governing the problem \citep{Szebehely1967}.  The equations of motion in 
the non-rotating coordinate system are obtained by taking $\mu \rightarrow 
0$ in Eqs (\ref{S6eq3a}) through (\ref{S6eq3c}), and in the synodic coordinate 
system by applying the same limit to Eqs (\ref{S6eq4a}) through (\ref{S6eq4c}).   

An extensive discussion of the {\it circular} Hill problem and its specific applications to the Sun-Earth-Moon system is given by \cite{Szebehely1967}, 
who also presents families of periodic orbits, the libration points and the 
ZVCs.  Applications of the {\it circular} Hill problem to the 
Sun-Jupiter-asteroid system was considered by \cite{Bolotin2000}, who proved 
the existence of an infinite number of periodic and chaotic (almost collision) 
orbits in the synodic coordinate system.  Moreover, some periodic orbits were identified by \cite{Zagouras1985}, and \cite{Henon2003} extended his previous 
work \citep{Henon1974} on the {\it circular} Hill problem and discovered new 
families of double and triple-periodic symmetric orbits. 

\subsection{Poincar\'e's qualitative methods and chaos}

\cite{Poincare1892} considered an unstable periodic solution to the CR3BP 
(or in some specific cases to the circular Hill problem) and represented it 
by a closed curve.  He also accounted for a family of asymptotic solutions 
located on two asymptotic surfaces.  The closed curve and the two asymptotic 
surfaces were his geometrical representations of the CR3BP \citep{BarrowGreen1997}.  
Poincar\'e demonstrated that the equations describing these surfaces could be 
written as 
\begin{equation}
{{x_2} \over {x_1}} = f(y_1, y_2)\ ,
\label{S6eq10}
\end{equation}

\noindent
where the asymptotic series for the first and second surface are given by 
$x_1 = s_1 (y_1, y_2, \sqrt{\mu})$ and $x_2 = s_2 (y_1, y_2, \sqrt{\mu})$, respectively.  In addition, the series must satisfy  
\begin{equation}
{{\partial F} \over {\partial x_1}}\ {{\partial x_i} \over {\partial y_1}} + 
{{\partial F} \over {\partial x_2}}\ {{\partial x_i} \over {\partial y_2}} + 
{{\partial F} \over {\partial y_i}} = 0\,  
\label{S6eq11}
\end{equation}

\noindent
where $i = 1$ and $2$, and $F = F_0 + \mu F_1 + \mu F_2 + ...$, with 
$F_0 \neq F_0 (y)$ but $F_1$, $F_2$, ... being functions of both $x$ 
and $y$; see the text below Eq. (\ref{S3eq4}). 

\begin{figure}
\includegraphics[width=0.9\linewidth]{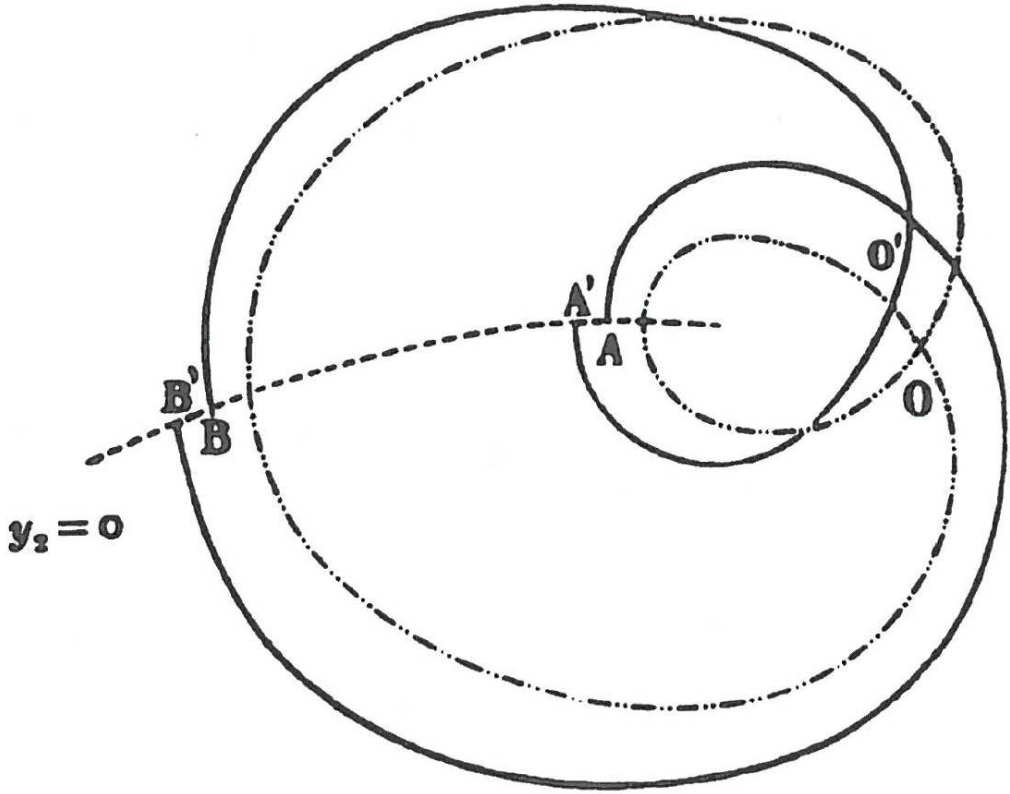}
\caption{Intersection of the asymptotic interfaces with the 
transverse section $y_1 = 0$ as explained in the main text. 
Reproduced after \cite{BarrowGreen1994}.  Copyright 1994 
Springer.}
\label{fig:Surfaces}
\end{figure}

Poincar\'e assumed that there were certain values $x_i^0$ for which 
${{\partial F} / {\partial x_i}}$ are commensurable, which allows 
writing $x_i = \Phi_i (y_1, y_2)$ with $F (\Phi_1, \Phi_2, y_1, y_2) 
= \tilde C$ satisfying Eq. (\ref{S6eq11}).  To integrate this equation, 
Poincar\'e considered $x_i = x_i^0 + (\mu x_i^1)^{1/2} + \mu x_i^2 + 
...$, and determined the coefficients $x_i^k$, with $k = 1$, $2$, ....  
The purpose of the second approximation was to evaluate an arbitrary 
number of coefficients of the above series, and to write the approximate 
equations for the asymptotic surfaces.  Finally, in the third approximation 
he constructed the intersection of the asymptotic interfaces with the 
transverse section $y_1 = 0$, and returned to his geometrical description, 
as depicted in Fig. \ref{fig:Surfaces}. 

In Fig. \ref{fig:Surfaces}, the closed curve represented by the 
dotted and dashed line corresponds to the generating (unstable) 
periodic orbit.  In addition, the curves $AO^{\prime}B^{\prime}$ 
and $A^{\prime}O^{\prime}B$ represent the asymptotic surfaces and 
their intersection with $y_1 = 0$, and the dashed line represents 
the curve $y_1 = y_2 = 0$.   Poincar\'e realized that the curves 
$AO^{\prime}B^{\prime}$ and $A^{\prime}O^{\prime}B$ could not be 
closed.  He considered possible trajectories that would simultaneously 
belong to both sides of the asymptotic surface and called them {\it 
doubly asymptotic} or {\it homoclinic orbits} \citep{Anderson1994}.  

Poincar\'e identified primary, secondary, tertiary, and quaternary 
homoclinic orbits and realized that there could be an infinite number 
of them leading to a {\it homoclinic tangle}, which appeared to be 
the first mathematical manifestation of chaos in the CR3BP; knowing 
how complicated the tangle could be, he did not attempt to draw it.  
Chaos has a sensitive dependence on initial conditions, and its 
existence in the CR3BP was confirmed by computational simulations 
of the CR3BP performed by \cite{Smith1993}, \cite{Koleman2012}, and 
many others.  Moreover, \cite{VelaArevalo2004} demonstrated that a 
method based on time-varying frequencies is a good diagnostic tool 
to distinguish chaotic trajectories from regular ones.  
      
Some of the modern computational methods use Poincar\'e sections to 
identify quasi-periodic and chaotic orbits.  \cite{Poincare1892} 
studied trajectories close to a periodic orbit and investigated 
their return to a line drawn across the orbit (now known as Poincar\'e 
section).  The behavior of trajectories returning to the line forms 
the first return map, or Poincar\'e's map, and he used it to reduce 
the dimension of phase space of the three-body problem by one.  He 
also showed that solutions could leave two equilibria and then return 
to them asymptotically, thus forming a {\it heteroclinic orbit}.  The 
above concepts originally introduced by Poincar\'e are now commonly 
used in modern theories of chaos \citep{Thompson1986,Hilborn1994,
Musielak2009}.

\section{The elliptic restricted three-body problem}

\subsection{Basic equations and integrals of motion}

Similar to the CR3BP, the elliptic restricted three-body problem 
(ER3BP) requires that the two objects (called the primaries) have 
their masses significantly larger than the third one ($M_1 >> M_3$ 
and $M_2 >> M_3$), and that the third mass has no gravitational 
influence on the primaries.  However, the main difference between 
the CR3BP and ER3BP is that, in the latter, motions of the primaries 
are along elliptical orbits around their center of mass, which has
important implications for the mathematical description of the system
as there is not a Jacobi's constant in the ER3BP \citep{Szebehely1967,
Marchal1990,Palacian2006a}.   

In general, the third mass can move in 3D space, a case is called the 
3D ER3BP.  However, if motions of the third body are restricted 
to the same plane as the primaries, we have the planar ER3BP.  
Typically, the governing equations for the 3D ER3BP are presented 
in the rotating-pulsating coordinates, which allowed for an elegant 
simplification of the problem \citep{Szebehely1967,Marchal1990}.  
In the following, we describe the governing equations of the 3D 
ER3BP, but then we make clear distinctions between the 3D and 
planar ER3BP when discussing some recently obtained results. 

Let $M_2$ move around $M_1$ on an elliptic orbit with eccentricity 
$e$, true anomaly $f$ and semimajor axis $a=1$, and let $\chi, \eta, 
\xi$ be the rotating-pulsating coordinates.  With $f$ being an 
independent variable, the set of governing equations describing 
the 3D ER3BP can be written \citep{Szebehely1967} in the 
following form:
\begin{equation}
{{\partial^2 \chi} \over {\partial f^2}} - 2 {{\partial \eta} \over
{\partial f}} = \omega_{\chi}\ , 
\label{S7eq1}
\end{equation}
\begin{equation}
{{\partial^2 \eta} \over {\partial f^2}} + 2 {{\partial \chi} \over
{\partial f}} = \omega_{\eta}\ , 
\label{S7eq2}
\end{equation}
and
\begin{equation}
{{\partial^2 \xi} \over {\partial f^2}} + {{\partial \xi} \over
{\partial f}} = \omega_{\xi}\ , 
\label{S7eq3}
\end{equation}

\noindent
where $\omega = \Omega / (1 + e \cos f)$ with $\Omega = (\chi^2 + \eta^2 
+ \xi^2) / 2 + (1 - \mu) / r_1 + \mu / r_2 + \mu (1 - \mu) / 2$, $r_1^2 = 
(\chi - \mu)^2 + \eta^2 + \xi^2$ and $r_1^2 = (\chi - \mu + 1)^2 + \eta^2 
+ \xi^2$.  For the {\it planar} ER3BP, it is necessary to take $\xi = 0$.

We now follow \cite{Llibre1990} and introduce $q_1 = - \chi + \mu$, $q_2 = 
- \eta$, $q_3 = \xi$, $p_1 = - \chi^{\prime} + \eta$, $p_2 = - \eta^{\prime} 
+ \mu$ and $p_3 = \xi^{\prime}$, where prime indicates $d / df$.  Then Eqs 
(\ref{S7eq1}) through (\ref{S7eq3}) can be written as 
\begin{equation}
{{dq_i} \over {df}} = {{\partial H} \over {\partial p_i}}\ , 
\hskip0.5in 
{{dp_i} \over {df}} = - {{\partial H} \over {\partial q_i}}\ ,  
\label{S7eq4}
\end{equation}

\noindent
where $i = 1$, $2$ and $3$, and $H$ is the time dependent Hamiltonian 
$H = H_0 + \mu H_1$ with 
\begin{equation}
H_0 = {1 \over 2} [ (p_1 + q_2)^2 + (p_2 - q_1)^2 + p_3^2 + q_3^2 ] 
- {{1} \over {1 + e \cos f}} \left [ {1 \over 2} \left (q_1^2 
+ q_2^2 + q_3^2 \right ) + {1 \over {r_1}} \right ]\ ,  
\label{S7eq5}
\end{equation}
and 
\begin{equation}
H_1 = {{1} \over {1 + e \cos f}} \left ( q_1^2 + {1 \over {r_1}} - 
{1 \over {r_2}} - {1 \over 2} \right )\ .
\label{S7eq6}
\end{equation}

\noindent
Since $H$ is an explicit function of $f$, it is not an integral of motion 
for the ER3BP; however, $H$ is an integral of motion for the CR3BP for 
which $e = 0$.  The planar ER3BP requires that $q_3 = p_3 = 0$. 

We may now apply the procedure of deriving the Jacobi integral for 
the CR3BP (see Section 6.3) to Eqs (\ref{S7eq1}) through (\ref{S7eq3}), 
which are the ER3BP governing equations.  The procedure requires 
multiplying Eqs (\ref{S7eq1}), (\ref{S7eq2}), and (\ref{S7eq3}) by 
$\chi^{\prime}$, $\eta^{\prime}$, and $\xi^{\prime}$, respectively, 
adding them and integrating.  The result is
\begin{equation}
\left ( {{\partial \chi} \over {\partial f}} \right )^2 + 
\left ( {{\partial \eta} \over {\partial f}} \right )^2 + 
\left ( {{\partial \xi} \over {\partial f}} \right )^2 =
2 \int ( \omega_{\chi} d \chi + \omega_{\eta} d \eta +
\omega_{\xi} d \xi )\ ,
\label{S7eq7}
\end{equation}

\noindent
which is equivalent to the Jacobi integral in the CR3BP; however, 
the difference is that the RHS of the above equation is not constant.  
Thus,  the Jacobi integral for the ER3BP is not a constant of motion.  

There are other constants of motions for the ER3BP.  As originally 
pointed out by \cite{Contopoulos1967}, two integrals of motion for 
the planar ER3BP exist if the system is considered in rotating 
coordinates with their $x$-axis passing through the primaries.  The 
results were generalized to the 3D ER3BP by \cite{Sarris1982}, 
who found three integrals of motion for a small eccentricity of the 
relative orbit of the primaries, and for a small distance of the 
third body from one of the primaries, and demonstrated that they 
depended periodically on time.

\subsection{Periodic solutions, libration points and Lie series solutions}

Periodic orbits showing chains of collision orbits of the original Kepler 
problem were first studied by \cite{Poincare1892}, who named them the 
{\it second species solutions}.  Searches for such orbits in the 3D ER3BP 
were performed by \cite{Gomez1991a}, \cite{Gomez1991b}, \cite{Bertotti1991}, 
\cite{Bolotin2000}, \cite{Bolotin2005} and \cite{Palacian2006b}.  In 
Bertotti's work, a variational method designed to find such periodic orbits 
was developed.  \cite{Bolotin2005} considered the planar ER3BP with a small 
mass ratio and eccentricity.  Then he proved the existence of many second 
species periodic orbits.  The fact that there are ejection-collision orbits 
in the 3D ER3BP was demonstrated by \cite{Llibre1990}.  Moreover, the 
effectiveness of the phenomenon of the gravitational capture of the third 
body by the primaries in the 3D ER3BP with small $\mu$ was discussed by 
\cite{Mako2004}.   

\cite{Sinclair1970} identified and classified various families of periodic 
orbits in the commensurable planar ER3BP, where two bodies are close 
to commensurability in their mean motions taking place in the same plane 
around the central body.  Moreover, \cite{Hadjidemetriou1992} and 
\cite{Haghighipour2003} discovered new families of stable 3:1 and 1:2 
periodic orbits in the planar ER3BP.   

In addition to the existence of periodic orbits in the 3D and planar 
ER3BP, there are also the equilibrium point solutions or the libration 
points in both problems.  The conditions that define the libration points 
are obtained from Eqs (\ref{S7eq1}) through (\ref{S7eq3}), namely, 
$\chi^{\prime \prime} = \eta^{\prime \prime} = \xi^{\prime \prime} = 
\chi^{\prime} = \eta^{\prime} = \xi^{\prime} = 0$ and $\partial \Omega 
/ \partial \chi = \partial \Omega / \partial \eta = \partial \Omega / 
\partial \xi = 0$.  Using these conditions, it is easy to show that there 
are five libration points in the ER3BP (similar to the CR3BP), and that 
three of them are collinear and two are triangular or equilateral 
\citep{Danby1964,Bennett1965}.  Motions around these libration points 
were investigated by \cite{Szebehely1967}, who demonstrated that the 
triangular points pulsate together with their own coordinate systems, 
and by \cite{Choudhry1977}, who showed that the coordinates of the 
libration points depend explicitly on time.         

A solution to the planar ER3BP was constructed by \cite{Delva1984}, 
who used the method of Lie series that is another form of the Taylor 
series \citep {Schneider1979}.  The ER3BP in the rotating-pulsating 
coordinate system was considered and the Lie operator for the motion 
of the third body was derived as a function of coordinates, velocities, 
and true anomaly of the primaries \citep{Delva1984}.  An analytical 
form of the Lie series was obtained and used to compute orbits; an 
extension of this method to the 3D ER3BP should be straightforward.
Numerical solutions to both the 3D and {\it planar} ER3BP were obtained 
and some specific examples are presented in the next section devoted 
to applications.

\subsection{Elliptic Hill problem}

We now briefly discuss the elliptic Hill problem and its potential 
applications to astrodynamics and extrasolar planetary systems.  As 
stated in Sections 1 and 6.7, in the three-body Hill problem $M_1 
>> M_2 >> M_3$.  If the orbit of $M_2$ around $M_1$ is elliptic, 
we then have the elliptic Hill problem \citep{Ichtiaroglou1980}.  
A few families of periodic orbits in the elliptic Hill problem 
were discovered by \cite{Ichtiaroglou1981}, who found all of them to 
be highly unstable.  In more recent work, \cite{Voyatzis2012} calculated 
a large set of families of periodic orbits in the {\it elliptic} Hill 
problem, determined their stability and applied the results to motions 
of a satellite around a planet.  \cite{Szenkovits2008} established a 
more accurate criterion of the Hill stability and used it to investigate 
stability of exoplanets in the systems: $\gamma$ Cephei Ab, Gliese 86 Ab, 
HD 41004 Ab, and HD 41004 Bb.  They concluded that the orbits of these 
exoplanets will remain stable, according to the Hill stability criterion, 
in the next few million years.

\section{Applications of the restricted three-body problems}

Among the possible solutions of the three-body problem, the simplification 
imposed in the restricted three-body problem has been the most practical,
and therefore many different astronomical systems have been studied within 
this framework.  We present applications of the restricted three-body 
problem with an emphasis on extrasolar planets and moons.  Since orbital 
dynamics plays a role in determining the likelihood of a planet being 
classified as habitable, residing in a region of space in which liquid water could persist on the surface of the planet, we provide additional discussion within that 
context. 

\subsection{The Earth-Moon system with a spacecraft}

The advent of the space exploration program presented a host of challenging 
three-body problems, and it established a new area of study in astrodynamics. 
The initial problem was that of solving the motion of artificial satellites, followed by the task of determining spacecraft orbital maneuvers required for missions to the Moon.

The orbital dynamics of artificial satellites and spacecraft is based 
on n-body theory, but it can be approximated as a three-body problem.  
Moreover, we distinguish between spacecraft trajectories and motion 
of artificial satellites (orbiters, space telescopes, space stations, 
communications satellites), as the solution for each case have different requirements and outcomes. The following are some interesting examples:

(i) For analysis of space missions, such a spacecraft moving between two 
planets, or from Earth to the Moon, we may apply the restricted three-body 
problem. However, space missions through the farther regions of the Solar 
System are rather complex and require that spacecraft trajectories be 
designed by decoupling an n-body system into several three-body systems. 
For example, by decoupling the Jovian moon n-body system into several 
three-body systems, it is possible to design a spacecraft orbit which 
follows a prescribed itinerary in its visit to Jupiter and its moons. In 
fact, the Jupiter-Ganymede-Europa-spacecraft system is approximated as 
two coupled planar three-body systems \citep{Gomez2001}.  Furthermore, for a spacecraft in transit from Earth to the outer planets, 
its flight may involve a gravity assist, which yields a four-body problem; 
once far away from the Earth, it reduces to a three-body problem. Further complications arise when spacecraft move in the vicinity of asteroids or 
other celestial objects with gravitational fields that are not sufficiently 
well-established.

(ii) For artificial Earth satellites, the problem of the orbit evolution 
is an instance of the four-body problem. The evolution of a satellite’s 
orbit with a high eccentricity, such as that of the Milniya communication satellites, is very different from that of a low-Earth orbit of low 
eccentricity (e.g., Hubble Space Telescope). In the latter case the most 
important perturbations are due to the oblateness of the Earth and the 
air drag. Both effects are small for a moderate to high eccentricity 
orbit with a large semi-major axis. But the gravitational influence 
of the Sun and the Moon becomes predominant. The Kepler Space Telescope 
orbits the Sun, which avoids the gravitational perturbations and torques 
inherent in an Earth orbit.

In general, the design of space missions requires analytical and numerical 
methods to solve planar circular restricted three-body problem (PCR3BP) 
and circular restricted three-body problem (CR3BP).  For example, when 
decomposing an n-body system into three-body systems that are not 
co-planar, such as the Earth-Sun-spacecraft and Earth-Moon-spacecraft 
systems, we need to consider three-dimensional configurations and solve 
circular restricted three-body problems (CR3BP).  The planar motion of a 
spacecraft relative to two other masses which orbit each other is very 
complex, with nonlinear effects due to the perturbation of the two 
bodies, for which no general analytical solution exists.  Therefore, 
the orbital analysis for such a spacecraft is based on sophisticated 
numerical methods. 

\subsection{The Sun-Earth-Moon system}

This particular solution of the three-body problem has influenced the 
theory of gravity since its inception by \cite{Newton1687}.  In comparison 
to the previous application concerning satellites, this approach relies 
on the Sun-Earth system as the two dominant bodies with the Moon as a 
test particle.  This was effectively the approach taken by Newton; 
however, he was not successful in determining stable solutions ($\sim8\%$ uncertainties).  The problem is that the mass of the Moon is comparable 
to that of the Earth within two orders of magnitude; it is also relatively 
close to the Earth, therefore, inducing reaction forces and tides that 
are not accounted for in the restricted three-body problem.  

Others have made progress in determining stability conditions for the 
Sun-Earth-Moon system.  \cite{Hill1877,Hill1878} developed a general 
stability criterion based upon the balance of gravitational forces on 
a test particle between the Earth and the Sun.  This criterion has 
been refined statistically through numerical simulation with estimates 
indicating a limit of $\sim3.2R_H$, where $R_H$ denotes the Hill Radius 
given by $R_H = \left({m_\oplus / 3 M_\odot}\right)^{1/3}a$ \citep
{Gladman1993}.  Limitations of this criterion in applications two 
extrasolar planetary systems were pointed out by \cite{Cuntz2009}.  
Recently, \cite{Satyal2013} used a method proposed by 
\cite{Szenkovits2008} that utilizes the Hill stability criterion; 
we discuss this approach in Section 8.6.

\subsection{The Sun-Jupiter system with an asteroid}

Beyond the Sun-Earth-Moon system, the next most studied restricted three-body 
problem is the Sun-Jupiter-asteroid system.  Since masses of asteroids are 
very small when compared to either the Sun or Jupiter, their description 
fits well within the framework of the restricted three-body problem.  Specific
applications include dynamics of asteroids within the Asteroid Belt, and the 
origin of the Kirkwood Gaps related to a depletion of the asteroid population in
some orbits due to resonances with Jupiter.  Resonances of asteroids with 
Jupiter have been studied by \cite{Nesvorny2002}, including possible three-body resonances in the Solar System.

As discussed in Section 6.2, the collinear Lagrange points are unstable. 
Thus, no population of asteroids in the vicinity of these points is expected.  
However, there is a population of asteroids with stable orbits around the 
$L_4$ and $L_5$ Lagrange points.  These populations are the Trojan asteroids,
and their stability has been investigated using the restricted three-body
formulation \citep{Beutler2005,Valtonen2006}. 

\subsection{A single star with giant and terrestrial exoplanets}

Since the confirmation of the first exoplanet 51 Pegasi around a solar-type 
star \citep{Mayor1995,Marcy1995}, we now know almost 1800 exoplanets, 
including more than 1000 exoplanets discovered by the NASA's {\it Kepler} 
space telescope; this number of discovered planets is growing everyday
{\footnote{http://exoplanet.eu/catalog/}}.  Prior to the launch of {\it 
Kepler}, most known exoplanets were Jupiter-analogs with respect to their 
sizes and masses.  A very small subset of these systems are now known to 
host terrestrial planets.  Thus, the restricted three-body problem 
has been used to investigate a planetary system with a single star, gas
giant, and terrestrial exoplanets.  

In studies performed by \cite{Noble2002}, broad regions of orbital 
stability were found for terrestrial exoplanets located in the 
habitable zone (HZ) of 51 Pegasi and other planetary systems.  More 
general stability limits for extra-solar planetary systems with two
exoplanets were established by \cite{Barnes2006}.  The HZ's for 
main-sequence stars of different spectral stars were originally 
established by \cite{Kasting1993}.  Prospects for the existence of
a terrestrial planet in stellar HZ with known giant exoplanets were
discussed by \cite{Jones2005}.  In more recent work, \cite{Kopparapu2010} identified 4 extra-solar planetary systems that can support a terrestrial 
planet in their HZ's.   

In the recent work of \cite{Goldreich2014}, resonances in the planar,
circular, restricted three-body problem with possibilities of semimajor 
axis migration and eccentricity damping result from planet-disk
interactions.  The authors showed that the migration and damping 
results in an equilibrium eccentricity, and that librations around 
this equilibrium are overstable.  Applications of these results to
exoplanets discovered by the {\it Kepler} satellite demonstrated that
most of the observed planets are overstable, which leads to passage
through resonance.  A more realistic case of the planar, three-body
problem is also considered with two (inner and outer) exoplanets being 
in resonance, and the departure from exact resonance is discussed.  

The most prolific detection method before {\it Kepler} was the radial 
velocity method, which relies on shifts of lines within the spectra 
of the host star.  These shifts demonstrate the reflex motion of the 
star due to the gravitational force exerted by the unseen companion.  
Terrestrial exoplanets would induce shifts in the radial velocity of 
the host star on the order of centimeters per second, which is just 
beyond today's limits for this detection technique.  In terms of 
exoplanet detections, the photometric transit method has uncovered 
through the {\it Kepler} sample, that giant and terrestrial planets 
can exist within the same system with a larger fraction of planets 
being Neptune-sized or smaller \citep{Howard2012,Fressin2013,Santerne2013}.  
In such systems the restricted three-body problem can be used to 
constrain the likelihood of additional planets.

\subsection{A single star with a giant exoplanet and exomoon}
\label{sec:exomoon}
Another set of systems has been proposed for the Jupiter-like planets 
discovered within a host star's HZ \citep{Kasting1993,Kopparapu2010}.  
Giant exoplanets within this region present a major challenge to the 
search for terrestrial exoplanets, which could support life.  It has 
been shown that giant exoplanets near the HZ can induce perturbations 
upon the putative terrestrial exoplanet, thereby making the orbit of 
the smaller body unstable.  In order to keep the search for habitability meaningful, exomoons have been considered as possible abodes for life, 
which orbit giant exoplanets within the HZ.  Studies of such systems 
are based on the restricted three-body problem.  

The plausibility of a Jupiter-like exoplanet located in the habitable zone 
to host an Earth-mass moon has been considered \citep{Williams1997}.  
The formation of such a system may experience considerable difficulties 
\citep{Canup2006}, but the capture of a terrestrial planet still remains 
possible \citep{Kipping2009,Kipping2012,Kipping2013a,Kipping2013b}.  Other 
studies have used a parameter survey approach to determine the stability 
of possible exomoons; see applications to the system HD 23079 system 
\citep{Cuntz2013}.  Studies of such systems are based on the formulation 
of the restricted three-body problem.

\subsection{Binary stellar systems with a giant or terrestrial exoplanet}

One of the most extreme cases of the restricted three-body problem 
involves the study of exoplanets orbiting either one or both components 
of a binary star system, in which the more massive and less massive stars 
are called the primary and secondary, respectively.  The categorization 
of these systems has been referred to as satellite-like (S-type) or 
planetary-like (P-type), where the former contains a planet that orbits 
a single star and the latter orbits both stars \citep{Dvorak1982}; the
extent of HZ's in such systems have been recently established \citep
{Eggl2012,Kane2013,Kaltenegger2013,Haghighipour2013,Cuntz2014}.   Using 
this convention, many different systems have been 
studied based on the restricted three-body problem, with various methods 
including topology and chaos indicators.  Furthermore, planetary formation 
in these systems have been studied even before one P-type system (Kepler16) 
with the exoplanet had been rigorously confirmed by \cite{Doyle2011}.  As 
this field encompasses a separate research area \citep{PilatLohinger2010}, 
we now discuss only selected recent approaches.

\begin{figure}
\centering
\includegraphics[width=0.5\linewidth]{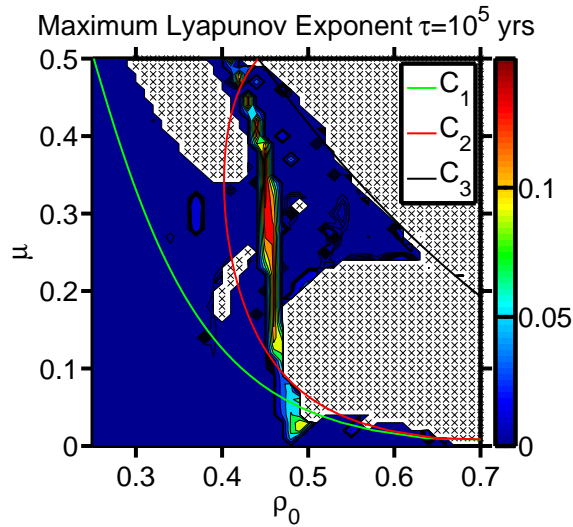}
\caption{Determination of stability of S-type planets using the method 
of Lyapunov exponents.  Regions of stability (blue) are indicated as 
well as regions of chaos (red).  Areas with a white background denote 
unstable regions on 0.1 Myr timescales.  Reproduced after \cite
{Quarles2011}.  Copyright 2011 Astronomy \& Astrophysics.}
\label{fig:stabContour}
\end{figure}

Using a topological method, \cite{Musielak2005} and \cite{Cuntz2007} 
determined stability criteria for the special case of binaries hosting 
exoplanets, which relied upon the Jacobi constant formalism in contrast 
to previous statistical approaches \citep{Holman1999,Dvorak1984}.  This 
approach allowed for the exploration of marginally stable systems, which
would have prematurely been characterized as stable using a parameter 
space consisting of the mass ratio $\mu = M_2 / (M_1 + M_2)$ and the 
initial distance ratio $\rho_o = R / D$, where $M_2 \leq M_1$, $R$ is 
the distance from the planet to $M_1$, and D is the initial separation 
of the stellar components.  

\cite{Eberle2010a} furthered this exploration of determining stability of 
planets in stellar binary systems using Hamilton's Hodograph \citep{Hamilton1847}.  This method demonstrated how the effective eccentricity of an exoplanet would 
change due to interactions with both stellar components.  \cite{Eberle2010b} 
explored the same parameter space and produced a map of general stability with contours of the effective eccentricity.  \cite{Quarles2011} used the parameter 
space along with the maximum Lyapunov exponent \citep{Lyapunov1907}, and found similar contours of stability with structures relating to orbital resonances 
and islands of instability.

Figure \ref{fig:stabContour} shows the parameter space with stability contours 
within a linear regime.  \cite{Quarles2011} showed that a 3:1 resonance can 
cause chaos within the circular restricted three body problem for a wide 
range of mass ratios.  Although these results have been mainly theoretical, 
they have been applied to the limited number of exoplanets found in stellar 
binary systems.  One such system, $\nu$ Oct, has a controversial discovery 
as the proposed distance ratio of the exoplanet lies within a region of 
instability (see Fig. 9).  This posed a problem for the confirmation of 
the exoplanet.  However, if one considers an exoplanet in retrograde 
orbit relative to the orbiting binary, then enlarged regions of stability 
arise, which would be consistent with the putative exoplanet \citep
{Eberle2010b,Quarles2012a,Gozdziewski2013}.

\begin{figure}
\centering
\begin{tabular}{cc}
\includegraphics[trim = 0mm 1mm 1mm 1mm, clip, width=0.4\linewidth]{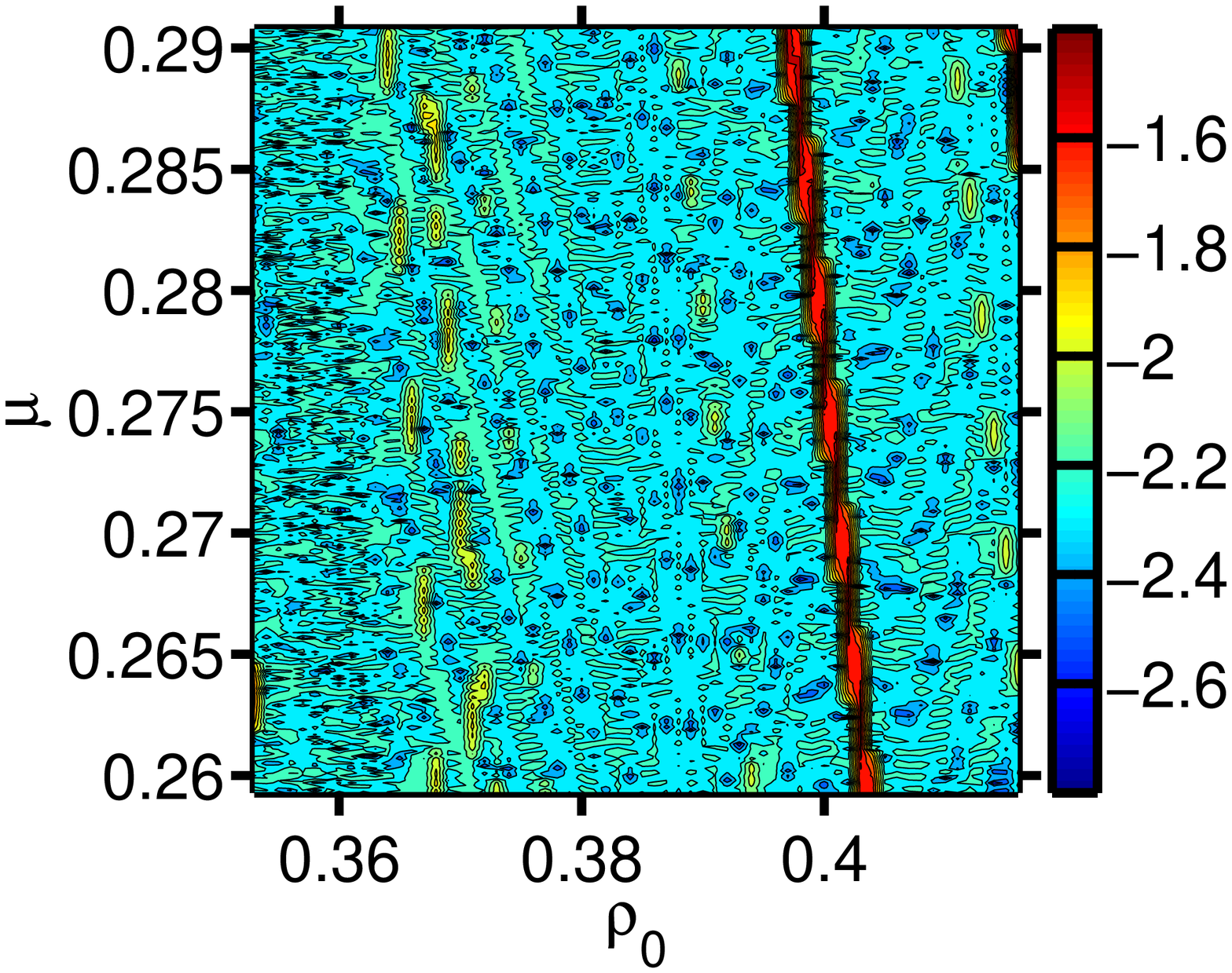}&
\includegraphics[trim = 0mm 1mm 1mm 1mm, clip, width=0.4\linewidth]{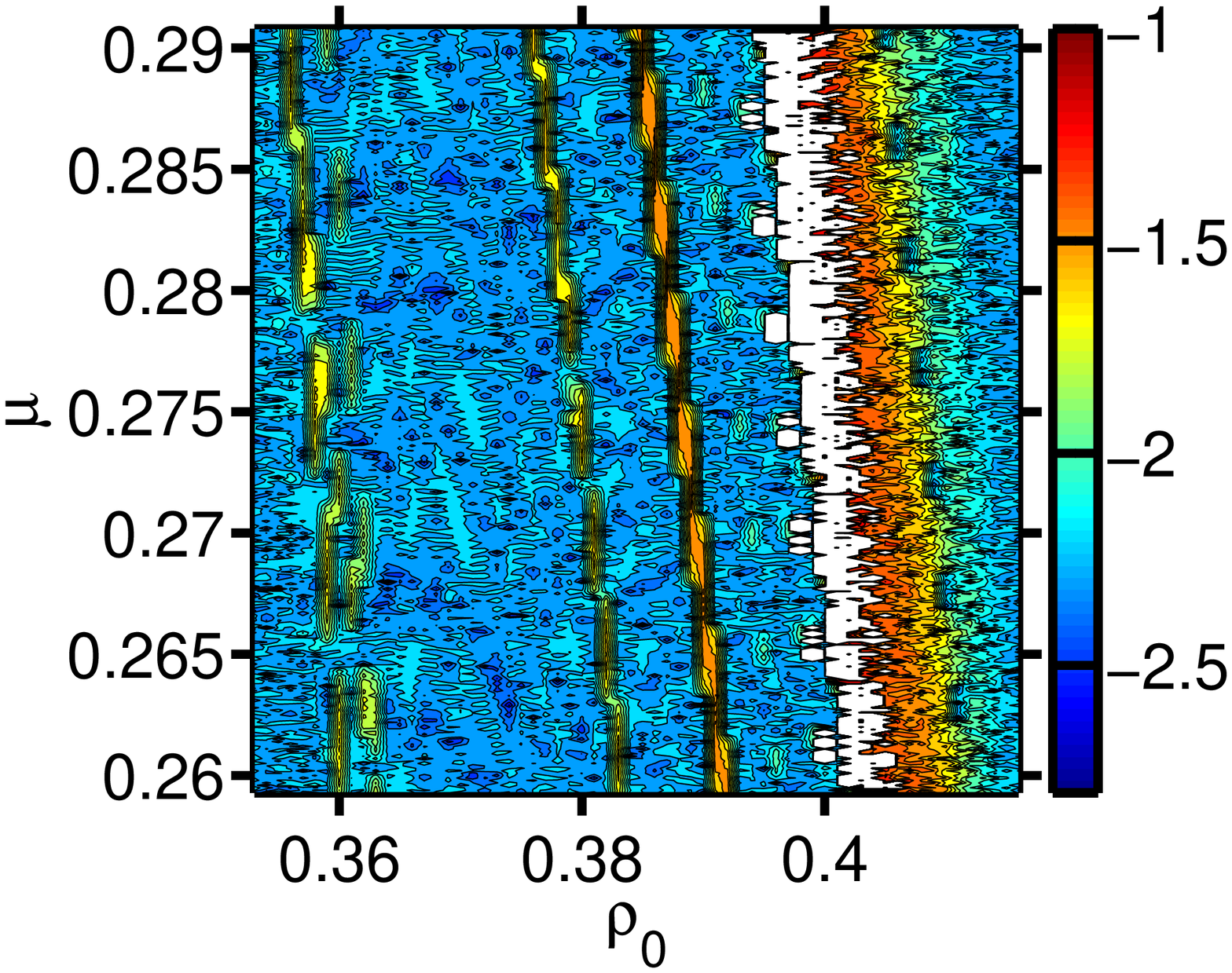}
\end{tabular}
\caption{Determination of stability of the planet in $\nu$ Octantis assuming 
a retrograde orbiting planet initially placed at apastron (left) and periastron 
(right).  Reproduced after \cite{Quarles2012a}.  Copyright 2012 Monthly Notices
of the Royal Astronomical Society.}
\label{fig:nuOct}
\end{figure}
 
The restricted three-body problem has also been applied to P-type or 
circumbinary exoplanets.  \cite{Doyle2011} announced the first circumbinary exoplanet in the Kepler-16 system.  This has been an important discovery for 
the theory behind the restricted three-body problem to be tested.  Soon after 
the discovery was announced, \cite{Quarles2012b} investigated cases when an 
Earth-mass exoplanet could be injected into the system; the studies of 
orbital stability of such an exoplanet could help in discovering habitable 
Earth-analogs in the Kepler-16 system.  

\begin{figure}
\centering
\includegraphics[width=0.5\linewidth]{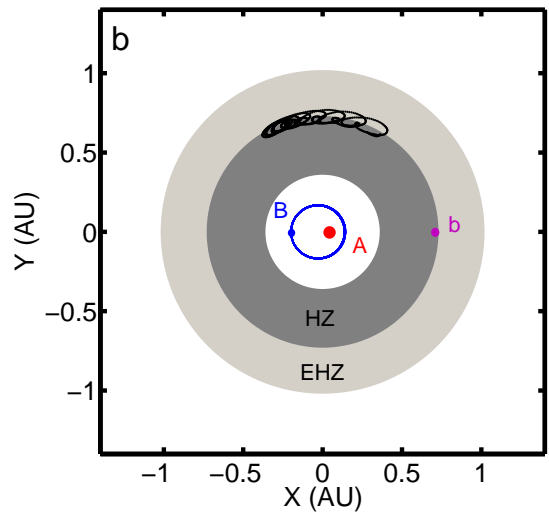}
\caption{Orbits of possible Trojan moons in Kepler-16 using a three-body approximation - see the main text for explanation.  Reproduced after 
\cite{Quarles2012b}.  Copyright 2012 The Astrophysical Journal.}
\label{fig:TrojanK16}
\end{figure}

Since injected terrestrial planets in the Kepler-16 system constitutes a 
restricted four-body system, \cite{Quarles2012b} reduced it to a restricted 
three-body problem in order to determine the possibility of the existence 
of Trojan moons in this system.  With the distance 
between the stellar hosts (denoted as A and B in Fig. 10), and the giant 
(Saturnian) exoplanet (denoted as b in Fig. 10) in the Kepler-16 system 
being relatively large, Earth-mass objects could remain in stable orbits 
around the approximated equilateral (L4 and L5) points in many different 
types of orbits.  Note that the giant planet is located at the border 
between the HZ and the so-called extended habitable zone (EHZ), which 
requires more extreme exoplanetary atmosphere with a significant 
back-warming \citep{Quarles2012b}.  Figure \ref{fig:TrojanK16} shows 
one of the orbits possible in this system; an exoplanet in this regime 
could be habitable, if it has a strong greenhouse effect \citep{Cuntz2014}.  However, either the formation or capture of such an exoplanet would be 
difficult to achieve within the environment of the circumbinary disk 
\citep{Meschiari2012,Paardekooper2012}.

\section{The relativistic three-body problem}

\subsection{The Newton and Einstein theories of gravity}

The three-body problem discussed in the previous sections is based on the 
Newtonian theory of gravity in which the bodies are assumed to be spherical 
and can be considered as point particles of given masses.  Since Newton's 
description of gravity is only a weak field approximation of Einstein's 
General Theory of Relativity (GTR), which is currently our best theory of 
gravity, now we want to discuss implications of using GTR on the three-body 
problem.  GTR replaces the flat space-time of Special Theory of Relativity 
by a 4D smooth, continuous, pseudo-Riemannian manifold endowed with the 
metric $ds^2 = g_{\mu \nu} (x) dx^{\mu} dx^{\nu}$, where $g_{\mu \nu}$ is 
the metric tensor to be determined by Einstein's field equations, $\mu$ and 
$\nu$ are 0, 1, 2 and 3, and the usual summation conventions are employed 
\citep{Hobson2006}.  According to GTR, the 4D space-time is curved by the 
presence of bodies and the resulting curvature is identified with gravity, 
which clearly shows that Newton's and Einstein's theories of gravity are significantly different.  Nevertheless, Einstein developed his GTR in such 
a way that in the limit of weak gravitational fields, Newton's description 
of gravity remains valid.

To account for relativistic effects on motions of astronomical bodies, 
relativistic celestial mechanics was formulated and its methods applied 
to many interesting astronomical problems, including the three-body 
problem \citep{Brumberg1972,Brumberg1991,Kopeikin2011}.  Since Einstein's 
field equations of GTR can only be solved for the simple one-body problem, 
and even the two-body problem cannot be solved rigorously in GTR, 
the equations of motion of the relativistic three bodies must be 
obtained with some approximations.  Typically, it is the so-called 
post-Newtonian approximation, which is a weak field and slow motion 
approximation, that is used to derive the post-Newtonian equations of 
motion of the three spherically symmetric bodies; the resulting equations 
depend not only on masses but also on physical parameters relevant to the 
internal structure of the bodies, such as elastic energy, moment of inertia, 
among others \citep{Brumberg1991}.  According to \cite{Kopeikin2011} and 
references therein, the presence of these parameters on the motions of 
the three bodies may actually be irrelevant.  Thus, some authors prefer 
to use the Einstein-Infeld-Hoffmann (EIH) equation of motion, which is 
valid for point-like masses with no internal structure 
\citep{Einstein1938,Infeld1957}.         

\subsection{The general relativistic three-body problem}

In the general relativistic three-body problem, there is no restriction 
on masses, except that they are finite.  The post-Newtonian 
equations of motion for this problem were obtained by \cite{Brumberg1972} 
and applied to different astronomical settings by \cite{Brumberg1991} as 
well as to the Solar System by \cite{Kopeikin2011}.  Hamiltonian and standard formalisms have also been formulated for the consideration of general relativistic effects within an n-body framework \citep{Saha1992,Saha1994,Kidder1995}.

The existence of the post-Newtonian collinear solution, which is a relativistic 
extension of the original Euler collinear solution (see Section 3.1 and Fig. 1), 
was shown by \cite{Yamada2010} who used the EIH equation of motion.  They also 
calculated the relativistic corrections to the spatial separation between the 
three bodies required for the solution to be valid.  The uniqueness of the 
relativistic collinear solution was demonstrated by \cite{Yamada2011}.

Moreover, a triangular solution that is a relativistic version of the Lagrange solution with an equilateral triangle joining the three bodies (see Section 
3.1 and Fig. 2) was originally obtained by \cite{Krefetz1967}.  More recently, 
\cite{Ichita2011} studied the post-Newtonian effects on this solution and 
showed that the solution satisfies the post-Newtonian equations of motion 
only when all three masses are equal.  Thereafter, it was demonstrated by 
\cite{Yamada2012} that the relativistic corrections to each side of the 
triangle must be added and, as a result, the relativistic triangle will 
not always be equilateral.  

Another interesting result is that the 8-type periodic solution for the 
general (non-relativistic) three-body problem with equal masses (see 
Section 3.2 and Fig. \ref{fig:Suvakov}a) does also exist in the general 
relativistic three-body problem.  Its first and second post-Newtonian 
orders were investigated numerically by \cite{Imai2007} and \cite{Lousto2008}, 
respectively, who demonstrated that the relativistic effects are responsible 
for only small changes in the shape of this 8-type solution.

It is also likely that relativistic versions of the 13 new periodic orbits 
found by \cite{Suvakov2013a} in the {\it planar general} (non-relativistic) 
three-body problem (see Section 3.2 and Fig. \ref{fig:Suvakov}) will be 
discovered and that relativistic corrections to the results presented in 
Fig. \ref{fig:Suvakov} will be found.  However, at the present time this 
idea still remains to be a conjecture.

\subsection{The restricted relativistic three-body problem}

In the restricted relativistic three-body problem (RR3BP), the two 
primaries have dominant masses and move around their center of mass; 
however, the third mass is very small and its gravitational influence 
on the primaries is negligible.  The post-Newtonian equations describing 
motions of the third body in the RR3BP were originally obtained by 
\cite{Brumberg1972}, who used a synodic (rotating) frame of reference 
with the origin at the center of mass and the primaries fixed.  

Let us consider a {\it planar} RR3BP, introduce the synodic coordinates 
$\upsilon$ and $\zeta$, and follow \cite{Douskos2002} and write the 
post-Newtonian equations of motion in the following form
\begin{equation}
\ddot \upsilon - 2n \dot \zeta = {{\partial U} \over {\partial \upsilon}} - 
{d \over {dt}} \left ( {{\partial U} \over {\partial \dot \upsilon}} \right )\ , 
\label{S9eq1a}
\end{equation}
and
\begin{equation}
\ddot \zeta + 2n \dot \upsilon = {{\partial U} \over {\partial \zeta}} - 
{d \over {dt}} \left ( {{\partial U} \over {\partial \dot \zeta}} \right )\ , 
\label{S9eq1b}
\end{equation}

\noindent
where `dot' represents the derivative with respect to time $t$, $n = 1 - 3 
[1 - \mu (1 - \mu) / 3] / 2 c^2$, with $c$ being the speed of light.  In 
addition, $U = U_c + U_r$ is the RR3BP potential function composed of the 
classical potential $U_c = r^2 / 2 + (1 - \mu) / r_1 + \mu / r_2$, with 
$r = \sqrt{\upsilon^2 + \zeta^2}$, $r_1 = \sqrt{(\upsilon + \mu)^2 + 
\zeta^2}$ and $r_2 = \sqrt{(\upsilon + \mu - 1)^2 + \zeta^2}$.  The 
relativistic corrections $U_r$ are given by    
\begin{align}
\centering
U_r &= - 6 r^2 c^2 \left [ 1 - {1 \over 3} \mu (1 - \mu) \right ] + {c^2 
\over 2} \left [ (\upsilon + \dot \zeta)^2 + (\zeta - \dot \upsilon)^2 
\right ]^2 \nonumber \\
&+\; 6 c^2 \left ( {{1 - \mu} \over {r_1}} + {{\mu} \over {r_2}} \right ) 
\left [ (\upsilon + \dot \zeta)^2 + (\zeta - \dot \upsilon)^2 \right ] 
- 2 c^2 \left ( {{1 - \mu} \over {r_1}} + {{\mu} \over {r_2}} \right )^2 
\nonumber \\
&-2 c^2 \mu (1 - \mu) \left [ {1 \over {r_1}} + \left ( {1 \over {r_1}} -
{1 \over {r_2}} \right ) (1 - 3 \mu - 7 \upsilon - 8 \dot \zeta) + \zeta^2 
\left ( {\mu \over {r_1^3}} + {{1 - \mu} \over {r_2^3}} \right ) \right ]\ .
\label{S9eq2}
\end{align}

The libration points for the RR3BP are obtained by substituting $\ddot 
\upsilon = \ddot \zeta = \dot \upsilon = \dot \zeta = 0$ in the above 
equations of motion \citep{Contopoulos1976,Bhatnagar1998}.  These 
libration points are the relativistic counterparts of the Lagrange 
collinear $L_1$, $L_2$ and $L_3$, and triangular $L_4$ and $L_5$ points.  
Linear stability of the relativistic collinear points was investigated 
by \cite{Ragos2000} and \cite{Douskos2002}, who showed that all these 
points were unstable, which is consistent with the results obtained 
for the non-relativistic collinear points. 

In the work by \cite{Bhatnagar1998}, linear stability of the relativistic 
triangular $L_4$ and $L_5$ points was studied and it was demonstrated 
that these points were unstable for the whole range $0 \leq \mu \leq 0.5$, 
despite the well-known fact that the non-relativistic $L_4$ and $L_5$ are 
stable for $\mu < \mu_0$, where $\mu_0 = 0.038521$ is the Routh critical 
mass ratio (see Section 6.2).  The problem was later revisited by \cite
{Douskos2002} and \cite{Ahmed2006}, who found that the relativistic 
triangular points are linearly stable in the range of mass rations 
$0 \leq \mu < \mu_r$, where $\mu_r = \mu_0 - 17 \sqrt{69} / 486 c^2$ 
\citep{Douskos2002}, and $\mu_r = 0.03840$ \citep{Ahmed2006}. 

Among different possible applications of the RR3BP as discussed by 
\cite{Brumberg1991} and \cite{Kopeikin2011}, let us also mention 
the calculation of the advance of Mercury's perihelion made by 
\cite{Maindl1984}, the computation of chaotic and non-chaotic 
trajectories in the Earth-Moon orbital system by \cite{Wanex2003}, 
the relativistic corrections to the Sun-Jupiter libration points 
computed by \cite{Yamada2010}, the analysis of stability of circular 
orbits in the Schwarzschild-de Sitter space-time performed by 
\cite{Palit2009}, and the investigation of the GTR effects in a 
coplanar, non-resonant planetary systems made by \cite{Migaszewski2009}.

\section{Summary and concluding remarks} 

We have reviewed the three-body problem in which three spherical 
(or point) masses interact with each other only through gravitational 
interactions described by Newton's theory of gravity, and no 
restrictions are imposed on the initial positions and velocities.  
We began with a historical overview of the problem, where a special 
emphasis was given to the difficulty of finding closed form solutions 
to the three-body problem due to its unpredictable behavior.  We gave 
detailed descriptions of the {\it general} and {\it restricted} 
(circular and elliptic) three-body problems, and described different 
analytical and numerical methods used to find solutions, perform 
stability analyses, and search for periodic orbits and resonances.  
We also discussed some interesting applications to astronomical 
systems and spaceflights.  In the previous section, we also 
presented the {\it general} and {\it restricted} relativistic 
three-body problem and discussed its astronomical applications. 

The three-body problem described in this paper may be considered as 
classical (or standard) because the bodies were assumed to be spherical 
objects or points of given masses as described by either Newton's theory 
of gravity (Sections 1-8), or by Einstein's theory of gravity (GTR) in 
its post-Newtonian approximation (Section 9).  However, there are studies 
of the effects of oblateness as well as Coriolis and centrifugal forces 
on the three-body problem \citep{Abouelmagd2013}, and the existence of 
the libration points in the restricted three-body problem with one or 
both primaries being oblate spheroids \citep{Markellos1996}, or both 
primaries being triaxial rigid bodies \citep{Sharma2001}.  Moreover, 
the existence of periodic orbits in the restricted three-body problem 
with one primary being an oblate body \citep{Mittal2009}, or with all 
objects being oblate spheroids and the primaries being radiation sources 
\citep{Abouelmagd2012}, have been investigated in the elliptic restricted 
three-body problem.  Similarly, the effects of photogravitational forces 
in such systems have also been studied \citep{Kumar2011,Singh2012}.
Note that we cite only selected papers in which references to previously 
published papers on these topics are given.

Also of interest is the so-called Chermnykh problem in which a point 
mass moves in the gravitational field produced by an uniformly rotating 
dumb-bell \citep{Gozdziewski1998,Gozdziewski2003}.  This problem 
generalizes two classical problems of celestial mechanics, namely, 
the two-fixed center problem and the restricted three-body problem, 
with the former applied to the semi-classical quantization of the 
molecular ion of hydrogen \citep{Gozdziewski1998}; however, any 
applications of the three-body problem outside of celestial mechanics 
and dynamical astronomy are outside of the scope of this paper.

We would like to point out that our reference list is only a small 
subset of all papers and books published on the three-body problem.  
The choice of references was determined by what we considered to be 
the most relevant to the topics covered in this paper; however, we 
do respect the fact that other authors may have different opinions 
in this matter.  Moreover, our reference list contains some papers 
with astronomical data where the results directly relate the 
applications of the general and restricted three-body problem to 
the Solar System, and newly discovered extrasolar planetary systems.

We hope that we presented a balanced view of the three-body problem, 
and that our choice of specific topics covered in this paper represents 
a good recognition of the previous and current research in the field.  
We also do sincerely hope that this review may serve as a guide to 
the major previous and current achievements in the field, and as an 
inspiration to scientists and students for opening new frontiers of 
research in this truly remarkable and very interesting problem.\\ \\ 

\noindent
{\bf Acknowledgments}

We want to thank many colleagues for sending us their published and 
unpublished papers, and for numerous discussions concerning the 
three-body problem.  We are especially indebted to M. Cuntz, J.L. Fry, 
K. Go\'zdziewski,  R. Hammer, J.J. Lissauer, D.E. Musielak, and A. Weiss 
for reading our manuscript and giving us valuable comments that help us 
to improve our presentation.  We borrowed several figures from already 
published papers, so we want to thank the original authors for giving 
us their permission to include these figures in our paper.  Z.E.M. 
acknowledges partial support of this work by the Alexander von Humboldt 
Foundation and by The University of Texas at Arlington through its 
Faculty Development Program. B.Q. acknowledges support by an appointment 
to the NASA Postdoctoral Program at Ames Research Center, administered 
by Oak Ridge Associated Universities through a contract with NASA.   

\bibliographystyle{plainnat}
\bibliography{references_sub}

\end{document}